\documentclass[AMA,Times1COL,compress]{WileyNJDv5}

\usepackage{subcaption}
\usepackage{bm}
\usepackage{xcolor}

\articletype{Research Article}%

\received{Date Month Year}
\revised{Date Month Year}
\accepted{Date Month Year}
\volume{00}
\copyyear{2025}
\startpage{1}

\begin{document}

\title{Nonlinear receding-horizon differential game for drone racing along a three-dimensional path}

\author[1]{Kijin Sung}
\author[1]{Kenta Hoshino}
\author[2]{Akihiko Honda}
\author[2]{Takeya Shima}
\author[1]{Toshiyuki Ohtsuka}

\authormark{Nonlinear receding-horizon differential game for drone racing along a three-dimensional path}

\address[1]{\orgdiv{Graduate School of Informatics}, \orgname{Kyoto University}, \orgaddress{\state{Kyoto}, \country{Japan}}} 
\address[2]{\orgdiv{Advanced Technology R\&D Center}, \orgname{Mitsubishi Electric Corporation}, \orgaddress{\state{Hyogo}, \country{Japan}}}

\corres{Toshiyuki Ohtsuka, Department of Informatics, Graduate School of Informatics, Kyoto University, Kyoto 606-8501, Japan. Email: ohtsuka@i.kyoto-u.ac.jp}


\fundingInfo{JSPS KAKENHI Grant Numbers JP22H01510 and JP23K22780.}

\abstract[Abstract]{Drone racing requires high-speed navigation through three-dimensional paths, posing significant challenges in control engineering. 
Existing control methods lack a feedback control framework that simultaneously addresses nonlinear drone dynamics and multi-agent competitive interactions, such as overtaking or obstructing opponents. 
To overcome this limitation, this study proposes a game-theoretic control framework, the nonlinear receding-horizon differential game (NRHDG), for competitive drone racing. 
NRHDG accounts explicitly for adversarial behavior by predicting and countering an opponent's worst-case behavior in real time. It extends standard nonlinear model predictive control (NMPC), which typically assumes a fixed opponent model. 
First, we develop a novel path-following formulation based on projection-point dynamics, eliminating the need for computationally expensive distance minimization during online control. 
Second, we propose a potential function that enables each drone to dynamically switch between overtaking and obstructing maneuvers, depending on the race situation. 
Third, we establish new performance metrics to evaluate NRHDG against NMPC across racing scenarios. 
Simulation results demonstrate that NRHDG outperforms NMPC in both overtaking and obstructing performance. 
Specifically, for randomly generated initial conditions and different levels of speed advantage for the rear-start drone, the 95\% confidence intervals for the arc-length-based mean performance differences excluded zero, indicating statistically significant advantages of NRHDG over NMPC in both overtaking and obstructing.}



\maketitle


\renewcommand\thefootnote{\fnsymbol{footnote}}
\setcounter{footnote}{1}

\section{Introduction} \label{sec:intro} 

Drone racing is an emerging field in robotics that requires drones to navigate three-dimensional paths at high speeds with precision\cite{race0}. 
These races pose significant control challenges, requiring trajectory adjustments under nonlinear dynamics. 
Although recent advances in trajectory optimization have enabled precise gate passages and energy-efficient maneuvers\cite{race1}, most existing methods have focused primarily on single-drone scenarios. 
Similarly, learning-based approaches can handle navigation through moving gates\cite{race2}; however, they typically neglect multi-agent competitive interactions such as overtaking or obstructing between the ego drone and its opponent. 

To address these limitations, several studies\cite{Spica2020,Wang2020,race3} have formulated drone racing as \emph{receding-horizon differential games} (RHDG). 
RHDG extends standard model predictive control (MPC) by predicting the opponent's worst-case actions over a finite horizon, thus optimizing the ego drone's strategy accordingly.
In particular, the nonlinear receding-horizon differential game (NRHDG) \cite{race3} accommodates fully nonlinear dynamics, rather than relying on simplified kinematic models\cite{Spica2020,Wang2020}. 
However, this approach still assumes fixed roles (i.e., overtaking or obstructing) for each drone, limiting the ability to switch strategies during a race. 
Moreover, relying on explicit path features, such as curvature and torsion, can increase computational costs and introduce numerical singularities.

In addition to competitive interactions, \emph{path-following control}\cite{Sujit2014,Rubi2020,Hung2023} is also essential for drone racing along three-dimensional paths. 
Path-following control aims to achieve timing-free progress along a given geometric path, distinguishing it from trajectory-tracking control that follows a time-dependent reference. 
Although path-following control is a crucial issue in control engineering, a unified and computationally efficient framework has not yet been fully established in the literature. 
Conventional path-following control methods involve iterative searches or approximations of an orthogonal projection (a \emph{projection point}) of a vehicle's position onto the path \cite{Hanson1995,Brito2019,Wang2022,Santos2023,Romero2022} or are limited to two-dimensional paths\cite{Altafini2002,Okajima2007}. 
Alternative approaches define reference points arbitrarily on the path, either through geometric relationships such as carrot chasing and virtual-target (pseudotarget) approaches\cite{Amundsen2023,Reinhardt2023,Xu2024,Degorre2024,Kumar2025}, or by introducing additional degrees of freedom via timing laws\cite{Faulwasser2009,Faulwasser2016}. 
Furthermore, there are other formulations of path-following control based on implicit function representations of paths \cite{Hladio2013,Chen2021,Itani2024} or vector fields\cite{Shivam2021}. However, arbitrary reference points do not represent the true distance from the path, and implicit function representations and vector fields are often difficult to construct and implement for general three-dimensional paths. 

This study integrates competitive interactions with role switching and a computationally efficient path-following formulation into the NRHDG framework. 
We first introduce a novel path-following model that integrates seamlessly with the drone's state equation without iterative minimization during online control. 
Subsequently, we extend the conventional NRHDG framework to facilitate dynamic role switching and reduce reliance on cumbersome path information. 
To achieve this, we introduce a potential function that allows each drone to flexibly switch between overtaking and obstructing modes in real time. 
Finally, we introduce systematic performance metrics to evaluate the advantage of NRHDG over standard nonlinear MPC (NMPC). The simulation results confirm that NRHDG outperforms NMPC in both overtaking and obstructing performance along a three-dimensional path.
The main contributions of this work can be summarized as follows.
\begin{enumerate}
    \item We derive a unified dynamical model for the orthogonal projection onto a three-dimensional path (Theorem \ref{thm:proj_ode} and equation (\ref{eq:stateeq_pathF}) in Section \ref{sec:model}). Unlike existing formulations that require distance minimization or frame-specific parameters, our model embeds directly into the drone's state equation and can be extended directly to smooth curves of arbitrary dimension. 
    The proposed dynamical model of the projection point provides a computationally efficient and unified formulation for path-following control, making it particularly suited for dynamic and complex environments such as drone racing. 
    \item We propose a custom smooth potential function for NMPC and NRHDG that enables real-time switching between overtaking and obstructing strategies, adapting to each drone's relative position (equation (\ref{eq:poten}) in Section \ref{sec:design}). 
    Its smoothness is well-suited to sensitivity-based numerical optimization involving fully nonlinear models, without introducing hard collision-avoidance constraints. 
    \item We introduce performance metrics to evaluate the effectiveness of overtaking and obstructing in drone racing (equations (\ref{eq:dsover}) and (\ref{eq:dsob}) in Section \ref{sec:design}). 
    We compare the performance metrics of different controllers in different races against the same opponent under identical initial conditions.
 Therefore, performance differences are attributed solely to the controllers. 
    Such comparisons were not considered in previous studies, in which controller performance was typically evaluated by swapping controllers between two drones.  
    Numerical simulations demonstrate that NRHDG outperforms NMPC (Section \ref{sec:sim}).
\end{enumerate}

The remainder of this paper is organized as follows. 
In Section 2, we derive the dynamical models of the drone and the projection of its position onto a three-dimensional path and state the control objectives. 
Section 3 provides an overview of NMPC and NRHDG, proposes objective functions for NMPC and NRHDG, including a potential function for role-switching, and then introduces performance metrics to evaluate the controllers. 
In Section 4, we conduct numerical simulations of races along a three-dimensional path and discuss the implications of these simulations. Finally, in Section 5, we state the conclusions of this study and discuss future work.

\section{Problem Formulation} \label{sec:model} 

\subsection{Drone dynamics} \label{subsec:model_drone}

\begin{figure}[tbp]
  \centering
  \includegraphics[height=0.18\textheight]{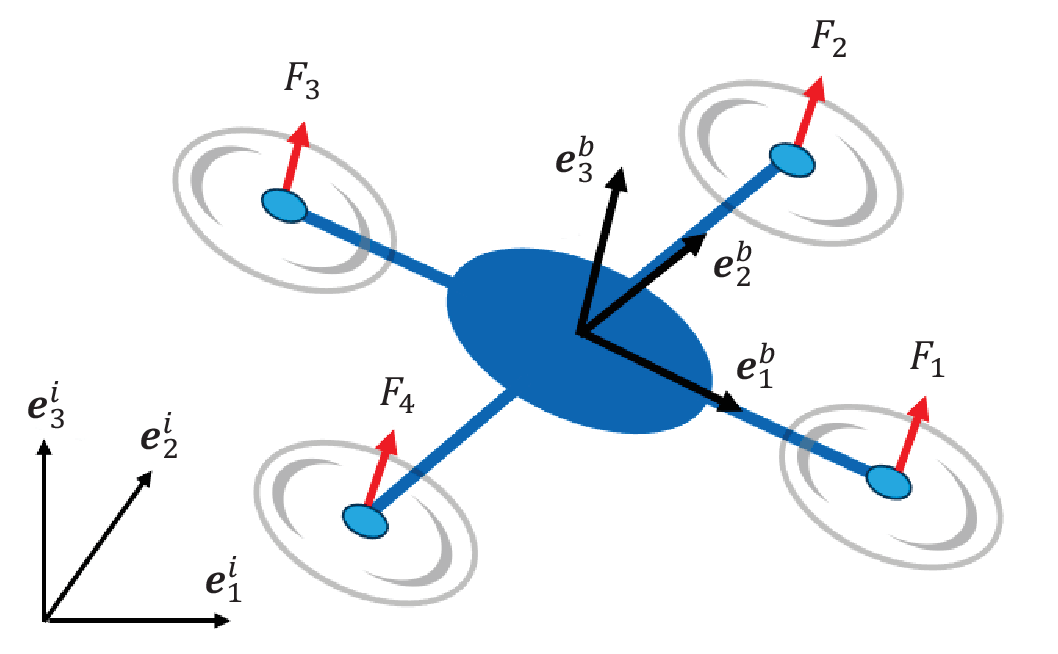}
  \caption{Overview of the inertial frame and body frame.} \label{fig:model}
\end{figure}

In this section, we present the dynamics of a drone, an augmented state equation for path following, and control objectives. 
We consider a quadrotor-type drone whose dynamical model is based on the model in reference\cite{Nonami2020}. 
The inertial frame is denoted by $(\bm{e}^i_1, \bm{e}^i_2, \bm{e}^i_3)$, and the body frame by $(\bm{e}^b_1, \bm{e}^b_2, \bm{e}^b_3)$ with origin at the drone's center of mass, as shown in Figure \ref{fig:model}, where the unit vector $\bm{e}^b_1$ points toward one of the four rotors, $\bm{e}^b_2$ points toward the next rotor counterclockwise, and $\bm{e}^b_3$ points upward. 
Attitude is parameterized by the unit quaternion\cite{Choset2005} $\bm{q}= (q_{0}\ q_{1}\ q_{2}\ q_{3})^{\rm T} \in \mathbb{R}^{4}$ with $\| \bm{q} \| = 1$. 
The rotation matrix $Q(\bm{q}) \in \mathbb{R}^{3 \times 3}$ from the body frame to the inertial frame is formed as follows:
\begin{equation}
Q(\bm{q}) = 
\left(\begin{array}{ccc}
    q_{0}^{2}+q_{1}^{2}-q_{2}^{2}-q_{3}^{2} & 2({q}_{1}{q}_{2}-{q}_{0}{q}_{3})& 2({q}_{0}{q}_{2}+{q}_{1}{q}_{3}) \\
    2(q_{1}q_{2}+q_{0}q_{3})& q_{0}^{2}-q_{1}^{2}+q_{2}^{2}-q_{3}^{2} &   2(q_{2}q_{3}-q_{0}q_{1})\\
    2(q_{1}q_{3}-q_{0}q_{2})& 2(q_{0}q_{1}+q_{2}q_{3})& q_{0}^{2}-q_{1}^{2}-q_{2}^{2}+q_{3}^{2} 
\end{array}\right).
\end{equation}

We assume that the external forces acting on the drone consist only of gravity in the $-\bm{e}^{i}_3$ direction and rotor thrusts in the $\bm{e}^{b}_{3}$ direction, and each rotor generates a reaction torque that is proportional to its thrust. 
In this study, we focus on the path following and competitive interactions of drones based on nonlinear dynamics, neglecting modeling errors and disturbances. 
Subsequently, the Newton-Euler equations describing the translational motion in the inertial frame and the rotational motion in the body frame of the drone are expressed as follows. 
\begin{equation}
  \left\{
    \begin{alignedat} {1}
      & m \ddot{\bm{p}}_d = Q_3(\bm{q}) \bm{e}_a^{\mathrm{T}} \bm{u}_{d} - mg \bm{e}^{i}_3 \\
      & J \dot{\bm{\omega}} + \bm{\omega} \times J \bm{\omega} = T \bm{u}_d,
    \end{alignedat}
  \right. \label{eq:N-E}
\end{equation}
where $\bm{p}_d = (x ~ y ~ z)^{\mathrm{T}} \in \mathbb{R}^{3}$ represents the drone's position in the inertial frame, $\bm{\omega} = (\omega_{1} ~ \omega_{2} ~ \omega_{3})^{\mathrm{T}} \in \mathbb{R}^{3}$ is the angular velocity vector in the body frame, and $\bm{u}_d = (F_1 ~ F_2 ~ F_3 ~ F_4)^{\mathrm{T}} \in \mathbb{R}^{4}$ is the vector of rotor thrusts. The symbols $m$ and $J \in \mathbb{R}^{3 \times 3}$ denote the mass and inertia matrix of the drone, respectively, $Q_3(\bm{q})$ denotes the third column of $Q(\bm{q})$, $\bm{e}_a = (1 \ 1 \ 1 \ 1)^{\mathrm{T}}$, and $T \in \mathbb{R}^{3 \times 4}$ is a matrix 
\begin{equation}
    T = \left(
    \begin{array}{cccc}
        0 & l & 0 & -l \\
        -l & 0 & l & 0 \\
        k & -k & k & -k \\
    \end{array}
    \right),
\end{equation}
where $l$ denotes the distance from the center of mass to each rotor and $k$ denotes the proportionality constant relating thrust to reaction torque. 
On the right-hand side of the first equation of (\ref{eq:N-E}), the first term consists of the three-dimensional vector $Q_3(\bm{q})$ multiplied by a scalar (inner product) $\bm{e}_a^{\mathrm{T}} \bm{u}_{d}$, and the second term consists of the three-dimensional vector $\bm{e}^{i}_3$ multiplied by a scalar $-mg$. 
The time derivative of the quaternion $\dot{\bm{q}}$ is given by
\begin{equation}
  \dot{\bm{q}} = \frac{1}{2} \varOmega(\bm{\omega}) \bm{q} \label{eq:qdot},
\end{equation}
where $\varOmega(\bm{\omega}) \in \mathbb{R}^{4 \times 4}$ is the following skew-symmetric matrix
\begin{equation}
  \varOmega(\bm{\omega}) = \left(
    \begin{array}{cccc}
      0 & -\omega_{1} & -\omega_{2} & -\omega_{3} \\
      \omega_{1} & 0 & \omega_{3} & -\omega_{2} \\
      \omega_{2} & -\omega_{3} &0 & \omega_{1} \\
      \omega_{3} & \omega_{2} & -\omega_{1} & 0
    \end{array}
  \right).
\end{equation}
We now define the 13-dimensional state vector $\bm{x}_d=( \bm{p}_d^{\textrm{T}} \ \dot{\bm{p}}_d^{\textrm{T}} \ \bm{\omega}^{\textrm{T}} \ \bm{q}^{\textrm{T}} )^{\textrm{T}}$. Combining  (\ref{eq:N-E}) and (\ref{eq:qdot}), the state equation for the drone is given as follows: 
\begin{equation}
\dfrac{d}{dt}
   \left(
    \begin{array}{c}
      \bm{p}_d \\ \dot{\bm{p}}_d \\ \bm{\omega} \\ \bm{q}
    \end{array}
  \right)
= \left(
    \begin{array}{c}
      \dot{\bm{p}}_d \\
      \frac{1}{m}Q_3(\bm{q}) \bm{e}_a^{\mathrm{T}} \bm{u}_d - g\bm{e}^{i}_{3} \\ -J^{-1}(\bm{\omega} \times J \bm{\omega} - T \bm{u}_d) \\
      \frac{1}{2} \varOmega(\bm{\omega}) \bm{q}
    \end{array}
  \right) \label{eq:stateeq_drone}.
\end{equation}

\subsection{Dynamics of projection point and arc length}

To enable efficient path-following control, we analyze the relationship between the drone's position and the path it should follow. 
By deriving the dynamics of a projection point and its associated arc length, we establish a foundation for efficient path-following control. 
We assume that the path is given as a curve parameterized by a \emph{path parameter} $\theta$ over an interval $\Theta \subset \mathbb{R}$. 
That is, the path is represented as the image $\bm{r}(\Theta) \subset \mathbb{R}^3$ of a mapping $\bm{r} : \Theta  \to \mathbb{R}^3$. 
We assume $\bm{r}$ is twice differentiable and regular, that is, $d\bm{r}(\theta)/d\theta \ne 0$ for all $\theta \in \Theta$. 
However, we do not assume that the mapping $\bm{r}$ is one-to-one globally, allowing the path to go through a point multiple times. 
Importantly, $\theta$ is not necessarily the arc length of the path, which allows us to have explicit representations of various paths. 

\begin{figure}[tbp]
  \centering
  \includegraphics[height=0.18\textheight]{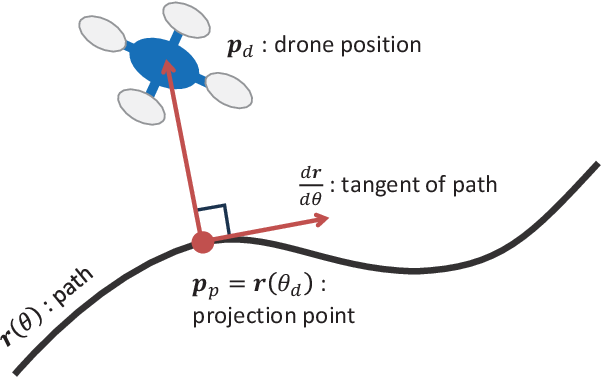}
  \caption{Relationship between drone and path.} \label{fig:path}
\end{figure}

We define the distance $d(\bm{p}_d)$ from the drone's position $\bm{p}_d \in \mathbb{R}^3$ to the path $\bm{r}(\Theta)$ by the minimum distance as 
\begin{equation}
    d(\bm{p}_d) = \inf_{\theta \in \Theta} \| \bm{r}(\theta) - \bm{p}_d \| . \label{eq:inf_d}
\end{equation}
If this infimum is attained at $\theta_d$ in the interior of $\Theta$, then the stationary condition 
\begin{equation}
     (\bm{r}(\theta_d) - \bm{p}_d)^{\mathrm{T}} \dfrac{d\bm{r}(\theta_d)}{d\theta} = 0 \label{eq:proj}
\end{equation}
holds. 
For $\theta_d$ satisfying (\ref{eq:proj}), we call $\bm{p}_p = \bm{r}(\theta_d)$ a \emph{projection point} of $\bm{p}_d$ because (\ref{eq:proj}) implies that $\bm{p}_p$ is an orthogonal projection of $\bm{p}_d$ onto the path, as shown in Figure \ref{fig:path}. 
We also define the \emph{signed arc length} $s(\theta_0,\theta_1)$ from $\bm{r}(\theta_0)$ to $\bm{r}(\theta_1)$ along the path as 
\begin{equation}
    s(\theta_0,\theta_1) = \int_{\theta_0}^{\theta_1} \left\| \dfrac{d\bm{r}(\theta)}{d\theta} \right\| d\theta 
\end{equation}
for $\theta_0, \theta_1 \in \Theta$, which is negative when $\theta_1 < \theta_0$. 
Since $d\bm{r}(\theta)/d\theta \ne 0$, $s(\theta_0,\theta_1)$ is strictly increasing with respect to $\theta_1$ for any fixed $\theta_0$. 
If the trajectory $\bm{p}_d(t)$ of the drone is differentiable in time $t \in [0,t_f]$ $(t_f > 0)$, we can derive ordinary differential equations for $\theta_d(t)$ and the corresponding arc length $\sigma(t) = s(\theta_{org},\theta_d(t))$, where $\theta_{org} \in \Theta$ is a fixed parameter chosen as the origin of the signed arc length coordinate.  

\begin{theorem}\label{thm:proj_ode}
Suppose the path $\bm{r}: \Theta \to \mathbb{R}^3$ is twice differentiable with $d\bm{r}(\theta)/d\theta \neq 0$ for any $\theta \in \Theta$, and the trajectory $\bm{p}_d: [0,t_f] \to \mathbb{R}^3$ of the drone is differentiable on $[0,t_f]$. 
If $\theta_d(t)$ is a solution of the following differential equation  
\begin{equation}
    \dot{\theta}_d(t) = \dfrac{\dot{\bm{p}}_{d}^{\mathrm{T}}(t) \dfrac{d\bm{r}(\theta_d(t))}{d\theta}}{\left\| \dfrac{d\bm{r}(\theta_d(t))}{d\theta} \right\|^2 + (\bm{r}(\theta_d(t)) - \bm{p}_d(t))^{\mathrm{T}}\dfrac{d^2 \bm{r}(\theta_d(t))}{d\theta^2} } 
 \label{eq:thetadot}
\end{equation}
with its initial value $\theta_d(0)$ satisfying (\ref{eq:proj}), and 
\begin{equation}
    \left\| \dfrac{d\bm{r}(\theta_d(t))}{d\theta} \right\|^2 + (\bm{r}(\theta_d(t)) - \bm{p}_d(t))^{\mathrm{T}} \dfrac{d^2 \bm{r}(\theta_d(t))}{d\theta^2} > 0  \label{eq:thetadot_denom}
\end{equation}
holds for all $t \in [0,t_f]$, then $\bm{p}_p(t) = \bm{r}(\theta_d(t))$ is a projection point corresponding to a local minimum distance to the drone's position $\bm{p}_d(t)$ for all $t \in [0,t_f]$. 
Moreover, the arc length $\sigma(t) = s(\theta_{org},\theta_d(t))$ of the projection point $\bm{p}_p(t)$ along the path satisfies
\begin{equation}
    \dot{\sigma}(t) = \left\| \dfrac{d\bm{r}(\theta_d(t))}{d\theta} \right\| \dot{\theta}_d(t)   \label{eq:sdot}
\end{equation}
for all $t \in [0,t_f]$ with the initial condition $\sigma(0) = s(\theta_{org},\theta_d(0))$. 
\end{theorem}

\begin{proof}
We obtain (\ref{eq:thetadot}) by differentiating (\ref{eq:proj}) with respect to time. 
In other words, (\ref{eq:thetadot}) implies that the time derivative of the left-hand side of (\ref{eq:proj}) is identically zero. 
Since the left-hand side of (\ref{eq:proj}) is equal to zero at $t=0$ by the assumption on the initial value $\theta_d(0)$, it remains zero for all $t \in [0,t_f]$ provided that (\ref{eq:thetadot}) holds. 
Moreover, (\ref{eq:thetadot_denom}) implies that the second-order derivative of $\| \bm{r}(\theta) - \bm{p}_d \|^2$ with respect to $\theta$ is always positive for $\theta = \theta_d(t)$ and $\bm{p}_d(t)$ for all $t \in [0,t_f]$. 
Therefore,  (\ref{eq:proj}) and (\ref{eq:thetadot_denom}) imply that $\theta_d(t)$ is a local minimizer of $\| \bm{r}(\theta) - \bm{p}_d \|^2$ for each $t \in [0,t_f]$. 
Note that the sufficient conditions for a local minimizer are valid even when $\theta_d$ is at the boundary of $\Theta$\cite{Nocedal1999}. 
Finally, by differentiating $\sigma(t) = s(\theta_{org},\theta_d(t))$ with respect to time, noting that $\theta_{org}$ is constant, we obtain (\ref{eq:sdot}) and observe that $\sigma(0) = s(\theta_{org},\theta_d(0))$ by definition. 
\end{proof}

\begin{remark}
Theorem \ref{thm:proj_ode} ensures that the path parameter $\theta_d(t)$ and the arc length $\sigma(t)$ of the projection point $\bm{p}_p(t)$ can be dynamically updated by integrating differential equations (\ref{eq:thetadot}) and (\ref{eq:sdot}), without requiring iterative searches or approximations of the projection point used in conventional methods\cite{Hanson1995,Brito2019,Wang2022,Santos2023,Romero2022}. 
Only the determination of the initial value $\theta_d(0)$ may require an iterative method. For a given initial position $\bm{p}_d(0)$, $\theta_d(0)$ can be obtained either by solving the minimization problem in (\ref{eq:inf_d}) or by solving (\ref{eq:proj}) and selecting a solution that satisfies (\ref{eq:thetadot_denom}) at $t=0$.
Once initialized, this differential-equation-based formulation simplifies path-following control by reducing computational overhead while maintaining generality across three-dimensional paths.
It also does not involve specific frames along the path, such as the Frenet-Serret frame, and it extends directly to curves of arbitrary dimension, in contrast to existing methods\cite{Altafini2002,Okajima2007}. 
\end{remark}

\begin{remark}
The singularity in the differential equations (\ref{eq:thetadot}) and (\ref{eq:sdot}) occurs when the denominator of (\ref{eq:thetadot}) vanishes, which is avoided if 
\begin{equation}
     \| \bm{r}(\theta_d(t)) - \bm{p}_d(t) \| \left \| \dfrac{d^2 \bm{r}(\theta_d(t))}{d\theta^2} \right\| < \left\| \dfrac{d\bm{r}(\theta_d(t))}{d\theta} \right\|^2   \label{eq:nonsingular_cond}
\end{equation}
holds, that is, if the deviation $\bm{r}(\theta_d(t)) - \bm{p}_d(t)$ or the second derivative $d^2\bm{r}(\theta_d(t))/d\theta^2$ of the path parameterization is sufficiently small. 
Moreover, Theorem \ref{thm:proj_ode} also guarantees that the projection point corresponds to a local minimum distance if the denominator of (\ref{eq:thetadot}) remains positive, which is the best possible guarantee without a global search. 
\end{remark}

\begin{remark}
If the path $\bm{r}$ is defined to satisfy $\| d\bm{r}(\theta)/d\theta \| = 1$, then $\dot{\sigma}(t) = \dot{\theta}_d(t)$ and $\sigma(t) = \theta_d(t) - \theta_{org}$ hold, and (\ref{eq:sdot}) can be omitted. 
However, it is often difficult to obtain an explicit arc-length parameterization for general three-dimensional paths. 
Therefore, (\ref{eq:sdot}) is useful in practical applications. 
\end{remark}

\subsection{Augmented state equation for path following}

To incorporate path-following dynamics into the drone's control framework, we augment the state vector $\bm{x}_d$ of the drone with the path parameter $\theta_d$ and the arc length $\sigma$. 
This extension enables seamless integration of path-following errors into the control design without requiring additional optimization steps. 
Specifically, since the right-hand sides of differential equations  (\ref{eq:thetadot}) and (\ref{eq:sdot}) depend on $\theta_d(t)$, $\bm{p}_d(t)$, and $\dot{\bm{p}}_d(t)$, we can incorporate them into the state equation of the drone. 
Subsequently, by augmenting the state vector as $\bar{\bm{x}}_d = ( \bm{p}_d^{\textrm{T}} \ \dot{\bm{p}}_d^{\textrm{T}} \ \bm{\omega}^{\textrm{T}} \ \bm{q}^{\textrm{T}} \ \theta_d \ \sigma ) ^{\textrm{T}} \in \mathbb{R}^{15}$, we obtain the augmented state equation for path following as follows: 
\begin{equation}
\dfrac{d}{dt}
   \left(
    \begin{array}{c}
      \bm{p}_d \\ \dot{\bm{p}}_d \\ \bm{\omega} \\ \bm{q} \\ \theta_d \\[2mm] \sigma
    \end{array}
  \right)
  = \left(
    \begin{array}{c}
      \dot{\bm{p}}_d \\
      \frac{1}{m}Q_3(\bm{q}) \bm{e}_a^{\mathrm{T}} \bm{u}_d - g\bm{e}^{i}_{3} \\ -J^{-1}(\bm{\omega} \times J \bm{\omega} - T \bm{u}_d) \\
      \frac{1}{2}\varOmega(\bm{\omega}) \bm{q} \\
      \frac{\dot{\bm{p}}_{d}^{\mathrm{T}} (d\bm{r}(\theta_d)/d\theta)}{\left\| d\bm{r}(\theta_d)/d\theta \right\|^2 + (\bm{r}(\theta_d) - \bm{p}_d)^{\mathrm{T}} (d^2 \bm{r}(\theta_d)/d\theta^2) }  \\
      \left\| \frac{d\bm{r}(\theta_d)}{d\theta} \right\| \frac{\dot{\bm{p}}_{d}^{\mathrm{T}} (d\bm{r}(\theta_d)/d\theta)}{\left\| d\bm{r}(\theta_d)/d\theta \right\|^2 + (\bm{r}(\theta_d) - \bm{p}_d)^{\mathrm{T}} (d^2 \bm{r}(\theta_d)/d\theta^2) }
    \end{array}
  \right) \label{eq:stateeq_pathF}.
\end{equation}
Equation (\ref{eq:stateeq_pathF}) defines the augmented state equation for the drone, including the dynamics of the path parameter $\theta_d$, which specifies the projection point, and the corresponding arc length $\sigma$. 
By embedding these dynamics directly into the state, the control system can account for path-following errors in real-time decision-making.
This projection-point dynamics model constitutes one of the main contributions of this study, providing a computationally efficient and unified path-following control formulation.

\subsection{Control objectives}

This study proposes feedback control strategies to achieve the following two primary objectives in competitive drone racing: 
\begin{enumerate}
    \item Efficient path following: Ensure that the ego drone progresses efficiently along the predefined three-dimensional path, minimizing deviation from the path while allowing flexibility for dynamic maneuvers. This flexibility avoids strict convergence to the path, which can hinder competitive behaviors such as overtaking or obstructing. 
    \item Dynamic overtaking and obstructing: Enable the ego drone to dynamically switch between overtaking a leading opponent and obstructing a trailing opponent, depending on the race scenario. These behaviors are designed to promote collision avoidance while maintaining competitive performance. 
\end{enumerate}

\section{Control methods and Objective Functions} \label{sec:control} \label{sec:design}

\subsection{Control system architecture}
\begin{figure}
    \centering
    \includegraphics[width=0.6\linewidth]{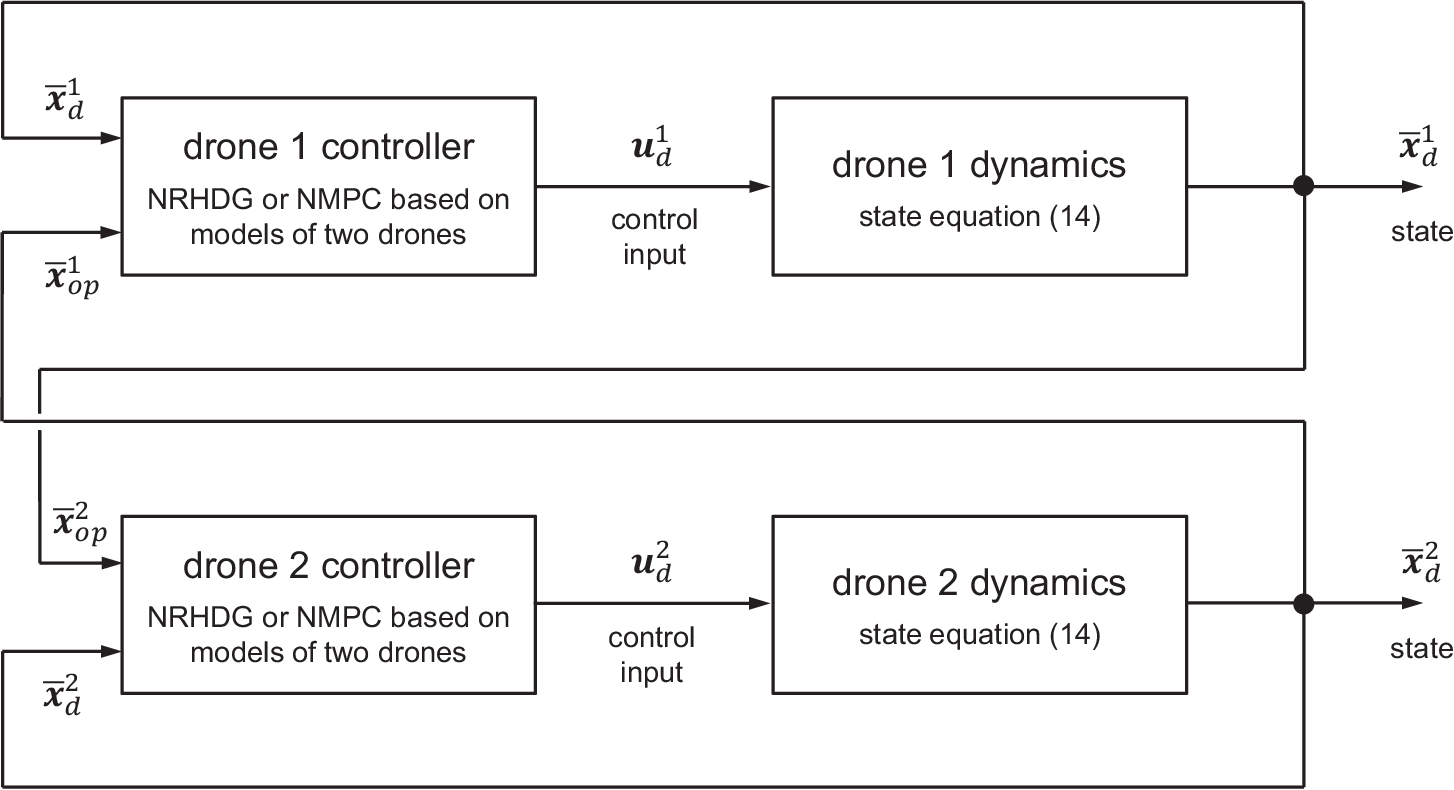}
    \caption{Block diagram of the feedback control system for competitive drone racing. Variables for each drone are indicated with superscripts. The state of drone 1, $\bar{\bm{x}}^1_d$, is used as the opponent's state $\bar{\bm{x}}^2_{op}$ in the controller of drone 2. In NRHDG, the objective function is defined by (\ref{rhdgef}), (\ref{eq:L_D}), and (\ref{eq:phi_D}), and the augmented state equation (\ref{eq:stateeq_pathF}) is used for predicting the motions of both the ego drone and the opponent. 
    In NMPC, the objective function is given by (\ref{eq:JJ}), (\ref{eq:raceL}), and (\ref{eq:racephi}), and the augmented state equation (\ref{eq:stateeq_pathF}) is used for predicting motion of the ego drone while the simplified state equation (\ref{eq:stateeq_MPC}) is used for predicting motion of the opponent.} 
    \label{fig:BlockDiagram}
\end{figure}

In competitive drone racing, the control system must simultaneously handle the nonlinear dynamics of the drone and potential adversarial behavior from an opponent. 
This section discusses two approaches: NMPC and NRHDG. 
First, we design an objective function for NMPC to achieve the control objectives. Subsequently, we modify it to define an objective function for NRHDG. 
The block diagram of the overall feedback control system for two drones is depicted in Figure \ref{fig:BlockDiagram}, where each drone optimizes its control input by solving an optimal control problem or a differential game problem over a finite prediction horizon based on the mathematical models of both the ego drone and the opponent. The state of the opponent is represented by $\bar{\bm{x}}_{op}$ and will be defined later. The variables for each drone are indicated with superscripts. For example, the state of drone 1, $\bar{\bm{x}}^1_d$, is used as the opponent's state $\bar{\bm{x}}^2_{op}$ in the controller of drone 2. Although NRHDG also predicts the opponent's control input for determining the best strategy, only the ego drone's computed optimal control input is applied in practice.

\subsection{Nonlinear model predictive control}

We first review NMPC, a widely used real-time optimization-based control approach for nonlinear systems, in a continuous-time setting\cite{jitsuyou}.
We consider a general dynamical system
\begin{equation}
	\dot{\bm{x}}_M(t) = \bm{f}_M(\bm{x}_M(t), \bm{u}(t), t), 
\end{equation}
where \(\bm{x}_M(t)\) is the state vector and \(\bm{u}(t)\) is the control input. 
To determine the control input $\bm{u}(t)$ at each time instant, NMPC solves a finite-horizon optimal control problem (OCP) to minimize the following objective function with a receding horizon $[t,t+T]$ $(T>0)$. 
\begin{equation}
	J_M[\bm{u}] = \varphi_M(\bm{x}_M(t+T)) + \int^{t+T}_t L_M(\bm{x}_M(\tau), \bm{u}(\tau),\tau)d\tau, \label{eq:JJ}
\end{equation}
where $\bm{x}_M(\tau)$ and $\bm{u}(\tau)$ $(t \leq \tau \leq t+T)$ represent the predicted state and control input over the horizon, respectively. They need not coincide with the actual system's state and control input in the future. 

At each time \(t\), the initial value of the optimal control input minimizing the objective function (\ref{eq:JJ}) over the horizon $[t,t+T]$ is used as the actual control input for that time. 
Subsequently, NMPC defines a state feedback control law because the control input $\bm{u}(t)$ depends on the current state $\bm{x}_M(t)$ of the system that is used as the initial state in the OCP over $[t,t+T]$. 
NMPC can handle various control problems if the OCP is solved numerically in real time. 
However, standard NMPC does not directly account for strategic multi-agent interactions in competitive scenarios such as drone racing. 
This limitation arises from the need to predict the opponent's behavior, which is inherently uncertain. Therefore, NMPC requires simplifying assumptions about the opponent's future trajectory, reducing its effectiveness in adversarial settings where opponents actively oppose the ego drone.

\subsection{Nonlinear receding-horizon differential game}

To explicitly model an adversarial opponent, we employ NRHDG, an extension of NMPC incorporating a differential game problem (DGP)\cite{dg0}. 
In what follows, we limit our discussion to a two-player zero-sum DGP, where one player's gain equals the other player's loss, making it a suitable model for competitive drone racing. 
We consider a dynamical system described by a state equation
\begin{eqnarray}
	\label{nonlinearsys2}
	\dot{\bm{x}}_D(t) = \bm{f}_D(\bm{x}_D(t), \bm{u}(t), \bm{v}(t), t),
\end{eqnarray}
where $\bm{x}_D(t)$ is the combined state vector of two players, \(\bm{u}(t)\) is the strategy (control input) for player $U$, and \(\bm{v}(t)\) is the strategy for player $V$. 
Player $U$ aims to minimize some objective function, while player $V$ aims to maximize it. 
We assume that both players know the current state of the game and the state equation governing the game.

In NRHDG, we consider an objective function with a receding horizon as follows:
\begin{equation}
	\label{rhdgef}
	J_D[\bm{u},\bm{v}] = \varphi_D(\bm{x}_D(t+T)) + \int^{t+T}_t L_D(\bm{x}_D(\tau), \bm{u}(\tau), \bm{v}(\tau),\tau)d\tau,
\end{equation}
which models the zero-sum nature of the game. 
If there exists a pair of strategies $(\bm{u}^0,\bm{v}^0)$ such that
\begin{equation}
J_D[\bm{u}^0,\bm{v}] \leq J_D[\bm{u}^0,\bm{v}^0] \leq J_D[\bm{u},\bm{v}^0]
\end{equation}
holds for all admissible strategies $\bm{u}$ and $\bm{v}$, $(\bm{u}^0,\bm{v}^0)$ is called a saddle-point solution. 
The saddle-point solution $(\bm{u}^0, \bm{v}^0)$ ensures that both players adopt optimal strategies, balancing minimization by player $U$ and maximization by player $V$.
In particular, player $U$ can achieve a lower value of the objective function if $V$ chooses a different strategy from $\bm{v}^0$. 
Therefore, $U$ can use the initial value of a saddle-point solution $\bm{u}^0$ for the DGP as the actual input to the system regardless of the control input of $V$, which also defines a state feedback control law for $U$. 
This saddle-point property makes NRHDG suitable for dynamic problems in competitive scenarios such as drone racing. 
Although the necessary conditions for a saddle-point solution in differential games are analogous to the stationary conditions for an OCP\cite{dg0}, numerical solution methods for NMPC are not necessarily applicable to NRHDG if they are tailored to minimization, such as descent methods that rely on line searches. 
However, there are some methods that are applicable to both NMPC and NRHDG\cite{race3,dg1,Nagata2023}.

\subsection{NMPC objective function}

\subsubsection{Path-following term}

Consider the augmented drone state $\bar{\bm{x}}_d$ in Section \ref{sec:model}, which includes position $\bm{p}_d$, velocity $\dot{\bm{p}}_d$, angular velocity $\bm{\omega}$, quaternion $\bm{q}$, path parameter $\theta_d$, and arc length $\sigma$. 
We define a stage cost for path following as follows: 
\begin{equation}
     L_{PF}(\bar{\bm{x}}_d,\bm{u}_d) = \sum_{i=1}^{3} a_{i}(p_{di} - r_{i}(\theta_d) )^{2}+\sum_{i=1}^{3}a_{i+3}\omega_{i}^{2} - a_{7} \sigma + \sum_{i=1}^{4} b(u_{di} - u_{ref})^{2},
     \label{eq:pathL}
\end{equation}
where \(u_{ref} = mg/4 \) represents the reference input when the drone is hovering. 
The parameters \(a_i \ (i = 1, \dots ,7)\) represent the state weights, while \(b\) serves as the input weight. 
The stage cost in (\ref{eq:pathL}) balances several competing objectives: the first sum penalizes deviations from the desired path. The second sum penalizes the drone's angular velocity to avoid excessive attitude motion. The third term maximizes the drone's progress along the path. The final sum suppresses large deviations in the control inputs from the reference input. 
The corresponding terminal cost 
\begin{equation}
    \varphi_{PF}(\bar{\bm{x}}_d)  = \sum_{i=1}^{3}a_{i}(p_{di} - r_{i}(\theta_d) )^{2}+\sum_{i=1}^{3}a_{i+3}\omega_{i}^{2} - a_{7} \sigma ,
    \label{eq:pathphi1}
\end{equation}
encourages the terminal state to align with the path-following objective. 
Since the path parameter $\theta_d$ and the arc length $\sigma$ of the projection point $\bm{p}_p = \bm{r}(\theta_d)$ are embedded in the state equation as state variables, the stage and terminal costs do not require a separate projection optimization problem or complicated coordinate transformation.

\subsubsection{Overtaking and obstructing term} \label{subsec:objective}

\begin{figure}[htbp]
    \centering
    \includegraphics[width=0.35\linewidth]{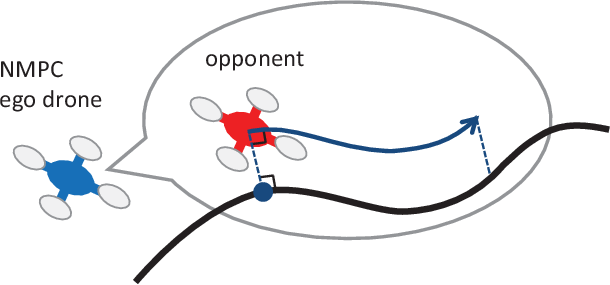}
    \caption{Prediction of opponent's motion in NMPC.}
    \label{fig:MPCpredict}
\end{figure}

To enable the ego drone to perform overtaking and obstructing maneuvers, we introduce a potential function that depends on both the ego drone's state $\bar{x}_d$ and the predicted state of the opponent.
For NMPC, we assume that the opponent moves at a constant path-parameter rate while maintaining a constant distance from the path (Figure \ref{fig:MPCpredict}). 
Specifically, the ego drone predicts the position $\bm{p}_{op}$ and path parameter $\theta_{op}$ of the opponent by a simplified state equation 
\begin{equation}
\dfrac{d}{dt} \left(
    \begin{array}{c}
      \bm{p}_{op} \\
      \theta_{op} \\ 
    \end{array}
  \right) =
  \left( \begin{array}{c}
      \lambda \frac{d{\bm{r}}(\theta_{op})}{d\theta} \\
      \lambda \\ 
    \end{array}
  \right) ,\label{eq:stateeq_MPC}
\end{equation}
where $\lambda$ denotes the constant path-parameter rate of the opponent. 
We denote the state vector of the simplified prediction model as $\bm{x}_{op} = (\bm{p}^{\textrm{T}}_{op} \ \theta_{op})^{\textrm{T}} \in \mathbb{R}^4$. 

\begin{figure}[htbp]
    \centering
    \includegraphics[width=0.35\linewidth]{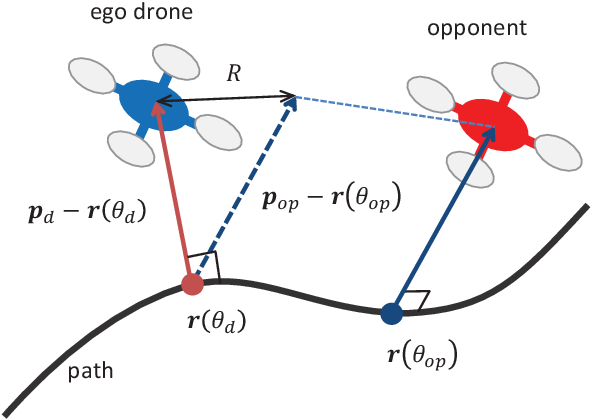}
    \caption{Difference between the two drones' deviations from the path.}
    \label{fig:dev_diff}
\end{figure}

\begin{figure}[tbp]
  \centering
  \includegraphics[width=0.4\linewidth]{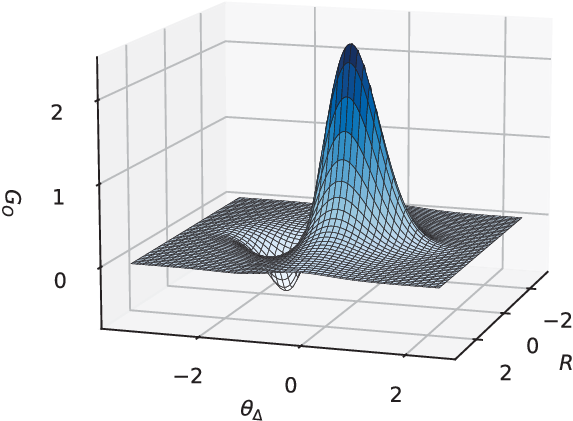}
  \caption{Potential function as a function of $\theta_\Delta$ and $R$. The plot is extended to negative values of $R$ solely for visualization because $G_O$ is even in $R$.} \label{fig:poten}
\end{figure}

Here, we define a potential function for overtaking and obstructing in NMPC as follows. 
\begin{align}
    G_{O}(\bar{\bm{x}}_d,\bm{p}_{op},\theta_{op}) &= \exp\left(-\left(\frac{\theta_{\Delta}-\delta_{1}}{\alpha}\right)^2\right) \tanh(\theta_{\Delta}-\delta_{2}) \frac{\beta}{1 + \gamma R^2}
    \label{eq:poten}, \\
    \theta_{\Delta} &= \theta_{op} - \theta_d, \quad R = \| (\bm{p}_{op} - \bm{r}(\theta_{op})) - (\bm{p}_d - \bm{r}(\theta_d)) \| , 
\end{align}
where $\theta_{\Delta}$ represents the difference between the path parameters of the ego drone and the opponent, $R$ is the norm of the difference between the deviations of the two drones from the path (Figure \ref{fig:dev_diff}), and $\alpha$, $\beta$, $\gamma$, $\delta_{1}$, and $\delta_{2}$ are constant shape parameters. 
The shape of this potential function is shown on the $\theta_{\Delta}$-$R$ plane in Figure \ref{fig:poten}, where the origin corresponds to the location of the ego drone. 
The potential function $G_O$ in (\ref{eq:poten}) enables switching between overtaking and obstructing strategies based on the relative position of the ego drone with respect to its opponent. 
When the ego drone follows the opponent, the ego drone should avoid and overtake the opponent. 
Therefore, we define the potential function such that it has its maximum at $R=0$ when $\theta_{\Delta}$ is larger than a threshold $\delta_2$. 
In this study, we use $R$, the norm of the difference between the deviations of the two drones from the path, rather than the distance $\| \bm{p}_{op} - \bm{p}_d \|$, because $R$ does not directly depend on the distance of the two drones along the path. 
That is, the ego drone does not need to slow down to keep the distance from the opponent along the path when it avoids and overtakes the opponent. 
However, when the ego drone precedes the opponent, the ego drone should obstruct the opponent to avoid being overtaken. 
To induce obstructing behavior, the potential function has its minimum at $R=0$ when $\theta_{\Delta} < \delta_2$. 
Since $R$ does not directly penalize the distance of the two drones along the path, the ego drone can maintain its speed when obstructing the opponent. 
This smooth potential function induces role switching and is one of the contributions of this study. Its smoothness is well-suited to sensitivity-based numerical optimization without introducing numerically cumbersome hard constraints for collision avoidance in existing approaches.

\subsubsection{Overall NMPC objective function} \label{subsec:NMPCformulation}

By combining the path-following term and the overtaking and obstructing term, we formulate NMPC for drone racing and define the overall objective function. The ego drone predicts its own future motion and that of the opponent using state equations (\ref{eq:stateeq_pathF}) and (\ref{eq:stateeq_MPC}), and the state vector of the entire system is defined as $\bm{x}_M =(\bar{\bm{x}}_d^\textrm{T} \ \bm{x}_{op}^\textrm{T})^\textrm{T} \in \mathbb{R}^{19}$. 
Subsequently, we define the stage cost and terminal cost of NMPC by combining the objective functions for path following and overtaking and obstructing, which enables the ego drone to balance path following with competitive behaviors, as follows.
\begin{align}
     &L_{M}(\bm{x}_M,\bm{u}_d) = L_{PF}(\bar{\bm{x}}_d,\bm{u}_d) + G_{O}(\bar{\bm{x}}_d,\bm{p}_{op},\theta_{op}),   \label{eq:raceL}\\
     &\varphi_{M}(\bm{x}_M)  = \varphi_{PF}(\bar{\bm{x}}_d) + G_{O}(\bar{\bm{x}}_d,\bm{p}_{op},\theta_{op}).     \label{eq:racephi}
\end{align}
These costs govern the behaviors of the ego drone depending on the race scenario represented by the states of the two drones. 
In particular, since the path parameters $\theta_d$ and $\theta_{op}$ are components of the state vectors determined by the state equations (\ref{eq:stateeq_pathF}) and (\ref{eq:stateeq_MPC}), there is no need for additional optimization or moving frames to determine the projection point along the path, which makes the proposed formulation advantageous over conventional formulations of path-following control.

\subsection{NRHDG objective function} \label{subsec:NRHDG}

In NRHDG, the ego drone minimizes an objective function (\ref{rhdgef}) while assuming that the opponent maximizes the same objective function. 
Furthermore, the ego drone assumes that the opponent is also governed by the same augmented state equation (\ref{eq:stateeq_pathF}). 
Unlike NMPC, which assumes a fixed prediction model for the opponent, NRHDG explicitly treats the opponent's control input as an adversarial decision variable. By incorporating a zero-sum differential game, NRHDG enables the ego drone to optimize its performance against an opponent that actively counters its actions.
We denote the state vector and the input vector of the opponent by $\bar{\bm{x}}_{op}$ and $\bm{u}_{op}$, respectively, which consist of the opponent's variables corresponding to those of $\bar{\bm{x}}_d$ and $\bm{u}_d$. 
We also denote the state vector of the entire system as $\bm{x}_D = ( \bar{\bm{x}}_d^\textrm{T} \ \bar{\bm{x}}_{op}^\textrm{T} )^\textrm{T}$. 
We now define the stage cost and the terminal cost for NRHDG as 
\begin{align}
    L_D(\bm{x}_D,\bm{u}_d,\bm{u}_{op}) &= L_{PF}(\bar{\bm{x}}_d,\bm{u}_d) + G_{O}(\bar{\bm{x}}_d,\bm{p}_{op},\theta_{op}) - L_{PF}(\bar{\bm{x}}_{op},\bm{u}_{op}) - G_{O}(\bar{\bm{x}}_{op},\bm{p}_d,\theta_d) , \label{eq:L_D}\\
    \varphi_D(\bm{x}_D) &= \varphi_{PF}(\bar{\bm{x}}_d) + G_{O}(\bar{\bm{x}}_d,\bm{p}_{op},\theta_{op}) - \varphi_{PF}(\bar{\bm{x}}_{op}) - G_{O}(\bar{\bm{x}}_{op},\bm{p}_d,\theta_d) . \label{eq:phi_D}
\end{align}
The stage cost $L_D$ in (\ref{eq:L_D}) includes terms for both the ego drone and the opponent, reflecting the adversarial nature of the interaction. 
The terminal cost $\varphi_D$ in (\ref{eq:phi_D}) evaluates the terminal state of the game, encouraging each drone's strategy to align with the race objectives. These costs induce a zero-sum strategic behavior in which the ego drone minimizes its cost while maximizing that of the opponent. 
At each time $t$, the ego drone determines its control input $\bm{u}_d$ by solving the NRHDG problem subject to the 30-dimensional state equation for $\bm{x}_D$.

\subsection{Performance metrics}

We assess the effectiveness of the proposed NRHDG against the baseline NMPC in competitive drone racing scenarios. 
In particular, we focus on the following two key aspects. 
\begin{enumerate}
    \item Overtaking performance: The ability of the following drone to maximize its progress while overtaking the preceding drone. 
    \item Obstructing performance: The ability of the preceding drone to minimize the following drone's progress while being overtaken.
\end{enumerate}
These aspects reflect both offensive (overtaking) and defensive (obstructing) capabilities of the controllers in adversarial races. 
In a race denoted by $\text{Race}(A,B)$ $(A, B \in \{\text{NMPC}, \text{NRHDG}\})$, the drone using controller $A$ starts ahead, while the drone using controller $B$ starts behind. We call $A$ the \emph{front-start controller} and $B$ the \emph{rear-start controller}, respectively. 

To evaluate the overtaking performance, we compare each pair of two races, $\text{Race}(A,\text{NMPC})$ and $\text{Race}(A,\text{NRHDG})$, for $A \in \{ \text{NMPC}, \text{NRHDG} \}$. 
This implies that we use the same front-start controller in both races and compare the progress of the rear-start controller. 
For a fair comparison of the controllers, the race settings, such as the initial lead and weight coefficients in the objective functions, are kept the same across the two races. 
Moreover, the race settings are chosen to ensure that the rear-start controller eventually overtakes the front-start opponent in all races, which enables us to make a quantitative comparison between different controllers in different races. 
A rear-start controller $B_1$ has better overtaking performance than the other rear-start controller $B_2$ if $B_1$ achieves more progress than $B_2$ while overtaking the same front-start controller $A$. 
In contrast, a front-start controller $A_1$ has better obstructing performance than the other front-start controller $A_2$ if the same rear-start controller $B$ achieves less progress against $A_1$ than against $A_2$. We define some quantities and performance metrics to quantify those performances in the following. 

In this study, we measure the progress of a drone by its arc length $\sigma(t)$ and use $\sigma^A_{(B)}(t)$ to denote the arc length of controller $A$ at time $t$ in $\text{Race}(A,B)$. 
Similarly, we use $\sigma^{(A)}_B(t)$ to denote the arc length of controller $B$ at time $t$ in $\text{Race}(A,B)$. 
In these symbols, the superscript denotes the front-start controller, the subscript denotes the rear-start controller, and the opponent is enclosed in parentheses. 
The same parameter $\theta_{org}$ is used as the origin of the arc length $\sigma(t)=s(\theta_{org},\theta_d(t))$ for both drones in all races under comparison, ensuring that their arc lengths are directly comparable. 
To specify the time to measure the progress of a drone, we define the \emph{overtaking time} $T^A_B$ in $\text{Race}(A,B)$. We first define the difference in arc length between the front-start controller $A$ and the rear-start controller $B$ as
\begin{align}
    \Delta \sigma^A_B (t) = \sigma^A_{(B)}(t) - \sigma^{(A)}_B (t). 
\end{align}
Since $\Delta \sigma^A_B(0) > 0$ holds at the initial time $t=0$, and $B$ overtakes $A$ when $\Delta \sigma^A_B(t) = 0$ holds for the first time, we define the overtaking time $T^A_B$ in $\text{Race}(A,B)$ as
\begin{align}
    T^A_B = \min \{ t \ge 0 : \Delta \sigma^A_B(t) = 0 \}.
\end{align}
We regard $T^A_B = \infty$ if $\Delta \sigma^A_B(t) > 0$ for all $t \ge 0$. 
Furthermore, to compare the progress between two races with the same front-start controller $A$ and different rear-start controllers, we define the \emph{maximum overtaking time $T_{\text{max}}^A$ for the front-start controller $A$} as
\begin{align}
    T^A_{\text{max}} = \max \{ T^A_{\text{NMPC}}, T^A_{\text{NRHDG}} \}. 
\end{align}
We also define the \emph{maximum overtaking time $T^{\text{max}}_B$ by the rear-start controller $B$} to compare the progress between two races with different front-start controllers and the same rear-start controller $B$ as
\begin{align}
    T^{\text{max}}_B = \max \{ T^{\text{NMPC}}_B, T^{\text{NRHDG}}_B \}. 
\end{align}

To compare overtaking performance between different controllers $B_1$ and $B_2$, we use their progress against the same front-start controller $A$ until both controllers overtake $A$ in $\text{Race}(A,B_1)$ and $\text{Race}(A,B_2)$, that is, $\sigma^{(A)}_{B_1}(T^A_{\text{max}})$ and $\sigma^{(A)}_{B_2}(T^A_{\text{max}})$. 
We use the maximum overtaking time for the same front-start controller $A$ here because a rear-start controller can delay overtaking to maximize its progress. 
If the relationship $\sigma^{(A)}_{B_1}(T^A_{\text{max}}) > \sigma^{(A)}_{B_2}(T^A_{\text{max}})$ holds in two races $\text{Race}(A,B_1)$ and $\text{Race}(A,B_2)$, $B_1$ advances farther than $B_2$ against the same opponent $A$ and has better overtaking performance than $B_2$. 
In the present setting, since NRHDG explicitly models the opponent's behavior and optimizes against the worst-case scenarios, we hypothesize that it will outperform NMPC. Therefore, we can expect NRHDG to have better overtaking performance than NMPC as follows. 
\begin{align}
     \sigma^{(A)}_{\text{NRHDG}}(T^A_{\text{max}}) > \sigma^{(A)}_{\text{NMPC}}(T^A_{\text{max}}), \quad A \in \{\text{NMPC},\text{NRHDG} \}, \label{eq:ADM} 
\end{align}
which means NRHDG achieves more progress than NMPC while overtaking the same controller $A$. For notational simplicity, we define the \emph{overtaking performance difference} $\Delta \sigma^{(A)}_{\text{over}}$ against a front-start controller $A$ as 
\begin{align}
    \Delta \sigma^{(A)}_{\text{over}} = \sigma^{(A)}_{\text{NRHDG}}(T^A_{\text{max}}) - \sigma^{(A)}_{\text{NMPC}}(T^A_{\text{max}}).  \label{eq:dsover}
\end{align}
If $\Delta \sigma^{(A)}_{\text{over}}$ is positive for $A \in \{\text{NMPC},\text{NRHDG} \}$ in numerical simulations, we can conclude that NRHDG has better overtaking performance than NMPC when overtaking $A$.

To compare obstructing performance between different controllers $A_1$ and $A_2$, we use the progress of the same rear-start controller $B$ until $B$ overtakes both $A_1$ and $A_2$ in $\text{Race}(A_1,B)$ and $\text{Race}(A_2,B)$, that is, $\sigma^{(A_1)}_B(T^{\text{max}}_B)$ and $\sigma^{(A_2)}_B(T^{\text{max}}_B)$. 
Here, we use the maximum overtaking time by the same rear-start controller $B$ because it can delay overtaking to maximize its progress. 
If the relationship $\sigma^{(A_1)}_B(T^{\text{max}}_B) < \sigma^{(A_2)}_B(T^{\text{max}}_B)$ holds in two races $\text{Race}(A_1,B)$ and $\text{Race}(A_2,B)$, controller $A_1$ slows the progress of the rear-start controller $B$ more effectively than $A_2$ and, therefore, has better obstructing performance than $A_2$. 
Here, we can expect NRHDG to have better obstructing performance than NMPC, and the following relationship holds. 
\begin{align}
    \sigma^{(\text{NRHDG})}_B(T^{\text{max}}_B) < \sigma^{(\text{NMPC})}_B(T^{\text{max}}_B), \quad B \in \{\text{NMPC},\text{NRHDG} \}, \label{eq:DMA}
\end{align}
which means NRHDG obstructs the progress of controller $B$ more effectively than NMPC. We also define the \emph{obstructing performance difference} $\Delta \sigma^{\text{ob}}_B$ against a rear-start controller $B$ as 
\begin{align}
    \Delta \sigma^{\text{ob}}_B = \sigma^{(\text{NRHDG})}_B(T^{\text{max}}_B) - \sigma^{(\text{NMPC})}_B(T^{\text{max}}_B).  \label{eq:dsob}
\end{align}
If $\Delta \sigma^{\text{ob}}_B$ is negative for $B \in \{\text{NMPC},\text{NRHDG} \}$ in numerical simulations, we can conclude that NRHDG has better obstructing performance than NMPC when obstructing $B$.

Since the performance differences in (\ref{eq:dsover}) and (\ref{eq:dsob}) are defined for different controllers in different races against the same opponent under identical initial conditions, they are attributed solely to the difference in the controllers. 
This was not the case in previous studies, where controller performance was typically evaluated by swapping the controllers between two drones (i.e., $\text{Race}(A,B)$ versus $\text{Race}(B,A)$), and is a key contribution of this study.

\section{Simulation} \label{sec:sim}

\subsection{Race setup}

\begin{figure}[htb]
  \centering
  \includegraphics[width=0.4\linewidth]{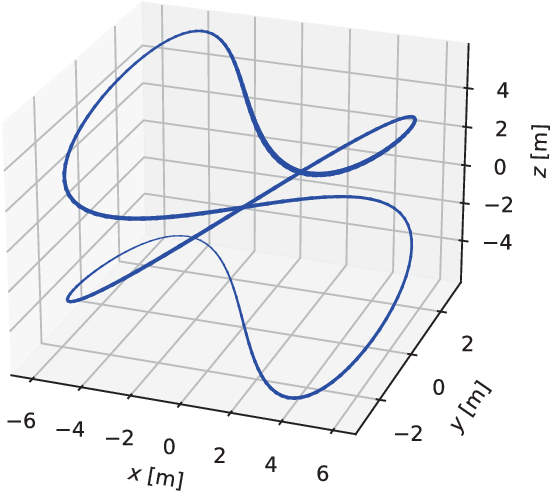}\\
 \caption{Overview of path.}\label{fig:course}
\end{figure}

We simulate races along a three-dimensional path: 
\begin{align}
  \bm{r}(\theta) =  \left(
    \begin{array}{ccc}
      6\text{sin} \theta & 3\text{sin} 2\theta & 6\text{sin} \frac{\theta}{2} 
    \end{array}
  \right)^\textrm{T},  \theta \in \mathbb{R} ,  \label{eq:course1}
\end{align}
depicted in Figure \ref{fig:course}. 
Although a single path is used across all simulations, it involves motion in all three spatial directions and exhibits diverse curvature and directional changes.
This provides a sufficiently challenging three-dimensional racing scenario for evaluating the relative performance of NRHDG and NMPC. 
The rear drone starts at $\bm{r}(0)$, and the front drone starts at $\bm{r}(1)$. 
In all simulations, we set $\theta_{org}=0$, which corresponds to the initial position of the rear-start drone on the path. Thus, the initial arc lengths of the rear and front drones are $0$ and $s(0,1)$, respectively. 
The physical parameters of the drones in the simulation are shown in Table \ref{tab:param}. These values are based on the Parrot MamboFly platform\cite{mambo}. 
The nominal values of the parameters in the objective functions of NMPC and NRHDG are shown in Table \ref{tab:pathomomi}. 
We assigned different input weights $b$ in the path-following term (\ref{eq:pathL}) to represent a speed advantage for the rear-start drone ($b=20$) and a slower response for the front-start drone ($b=40$). 

\begin{table}[tbp]
  \centering
  \caption{Parameters of drones.}
    \begin{tabular}{lll} \toprule
    Variable & Meaning & Value \\ \midrule
    $m$ & Mass of the drone & 0.063 kg \\ 
    $g$ & Gravitational acceleration & 9.81 m/$\mathrm{s^{2}}$ \\ 
    $l$ & Distance from the center of mass to the rotor & 0.0624 m \\ 
    $J$ & Inertia matrix & $\mathop{\mathrm{diag}}(J_{xx},J_{yy},J_{zz})$ \\
    $J_{xx}$ & Moment of inertia around the roll axis &
    $5.82857 \times 10^{-5}$ $\mathrm{kg \cdot m^{2}}$ \\ 
    $J_{yy}$ & Moment of inertia around the pitch axis & $7.16914 \times 10^{-5}$ $\mathrm{kg \cdot m^2}$ \\ 
    $J_{zz}$ & Moment of inertia around the yaw axis & $1 \times 10^{-4}$ $\mathrm{kg \cdot m^2}$ \\ 
    $k$ & Proportional constant between reaction torque and thrust & 0.0024 m \\ \bottomrule
    \end{tabular}
  \label{tab:param}
\end{table}

\begin{table}[tbp]
  \centering
  \caption{Nominal parameter values in the objective function.}
  \begin{tabular}{ll} \toprule
    Parameter & Value\\ \midrule
    $a_i \ (i=1, \ 2 , \ 3)$ & 1 \\ 
    $a_i \ (i=4, \ 5 , \ 6)$ & 0.1 \\ 
    $a_7 \  $ & 0.5 \\ 
    $b$ (front-start) & 40 \\
    $b$ (rear-start)  & 20 \\
    $\alpha $ & 1 \\  
    $\beta  $ & 4 \\ 
    $\gamma $ & 5 \\ 
    $\delta_{1} $ & $-0.5$ \\ 
    $\delta_{2} $ & $-1$ \\ 
    $\lambda$ & $1$ \\ \bottomrule
  \end{tabular}
  \label{tab:pathomomi}
\end{table}

We implemented NRHDG and NMPC with a continuation-based real-time optimization algorithm, C/GMRES\cite{Ohtsuka2004cgmres}, and its automatic code generation tool, AutoGenU for Jupyter\cite{Katayama2020autogen}\footnote{\url{https://ohtsukalab.github.io/autogenu-jupyter}}. 
C/GMRES computes a stationary solution to an optimal control problem without any line searches and is also applicable directly to NRHDG problems. 
AutoGenU for Jupyter generates C++ code and a Python package for updating the solution with C/GMRES. Subsequently, the generated code for NRHDG and NMPC can be used together in simulations of drone racing. 
We conducted numerical simulations on a PC (CPU: Core i9-12900 2.4 GHz, RAM: 16 GB, OS: Ubuntu 22.04.2 LTS running under WSL2 on Windows 11 Pro) to demonstrate the feasibility of real-time implementation. 
The simulation ran for 20 s, with a horizon length $T$ of 0.4 s and a control cycle of 1 ms. 
The average computation times per update were 0.8 ms for NRHDG and 0.5 ms for NMPC, both of which are within the control cycle.

\subsection{Results}

\subsubsection{Time histories and example overtaking scenario}

\begin{figure}[htbp]
    \centering
    \includegraphics[width=0.8\linewidth]{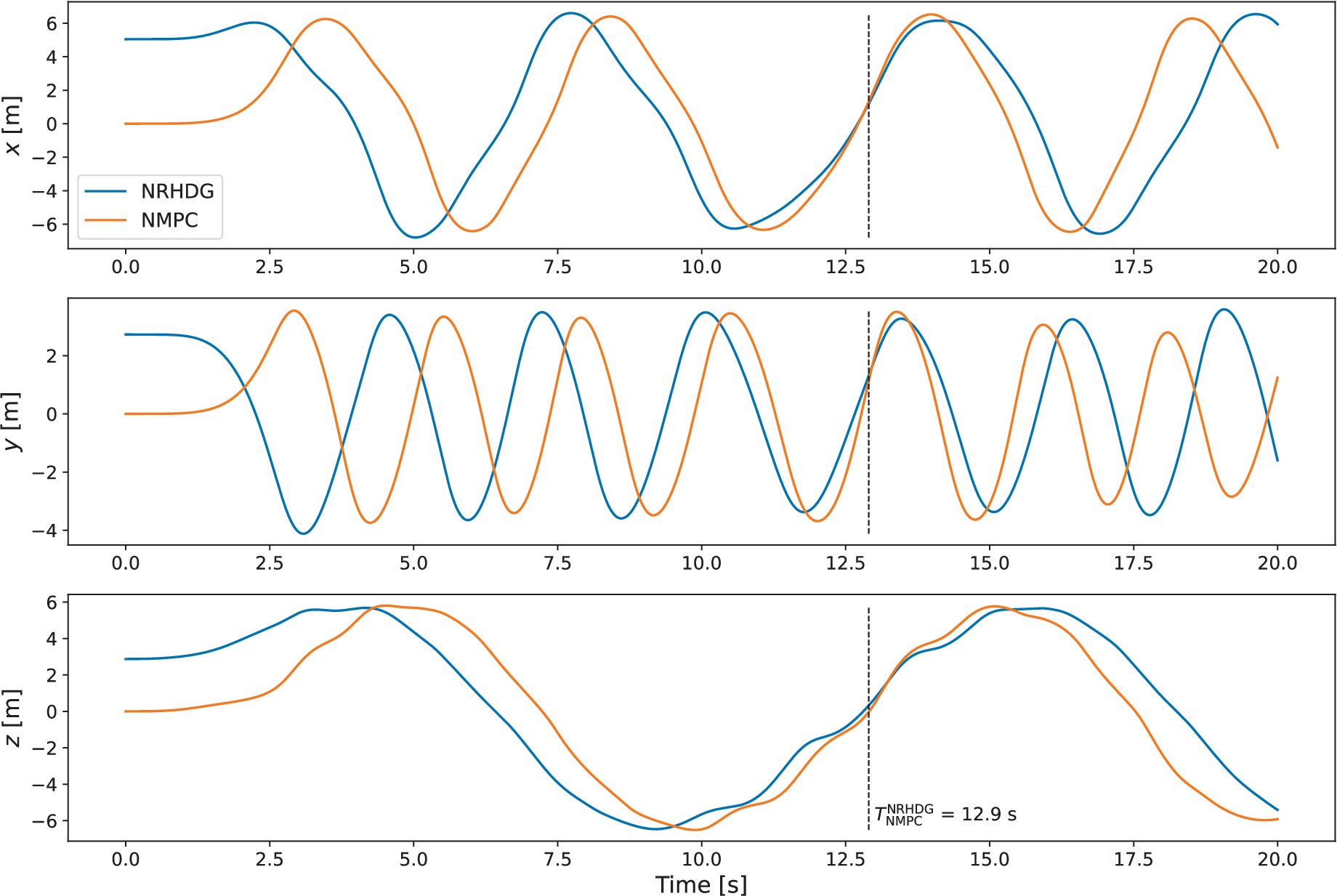}
    \caption{Time history of $\text{Race}(\text{NRHDG},\text{NMPC})$. Blue lines indicate the coordinates of the front-start NRHDG, and orange lines indicate those of the rear-start NMPC. The vertical dashed lines indicate the overtaking time $T^\text{NRHDG}_\text{NMPC}$. }
    \label{fig:timehis}
\end{figure}

\begin{figure}[htbp]
    \centering
    \includegraphics[width=0.8\linewidth]{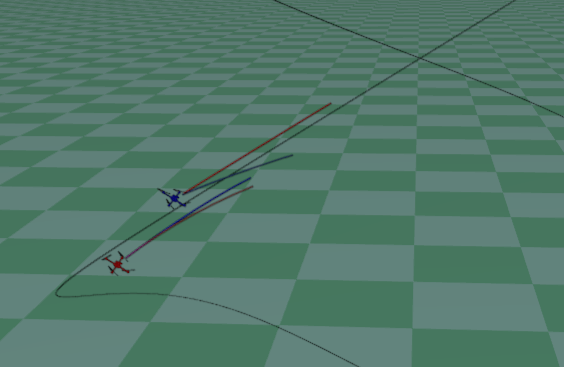}
    \caption{Snapshot of $\text{Race}(\text{NRHDG},\text{NMPC})$, where the blue drone (NRHDG) is obstructing the red drone (NMPC). The blue trajectories represent the predictions by the blue drone (NRHDG) of its future trajectory and that of the opponent. The red trajectories represent the predictions by the red drone (NMPC) of its future trajectory and that of the opponent. }
    \label{fig:racePic}
\end{figure}

Figure \ref{fig:timehis} shows a sample time history from $\text{Race}(\text{NRHDG},\text{NMPC})$, where NRHDG starts ahead and NMPC starts behind. 
The vertical dashed lines indicate the moment NMPC manages to overtake NRHDG. 
The altitudes ($z$) of the two drones oscillate as they attempt to obstruct or overtake one another. 
Figure \ref{fig:racePic} shows a snapshot of $\text{Race}(\text{NRHDG},\text{NMPC})$ around the overtaking time and compares the trajectories predicted by the two controllers over their respective prediction horizons. 
The trajectory colors indicate the controllers that generated the predictions: blue for NRHDG and red for NMPC.
NRHDG plans the blue drone's trajectory toward a position ahead of the red drone, while NMPC predicts the blue drone to move approximately parallel to the path. Conversely, NRHDG predicts the red drone to move approximately parallel to the path, while NMPC plans a larger avoidance maneuver for the red drone than that predicted by NRHDG. 
This indicates that NRHDG generates a less conservative prediction regarding the opponent's future behavior compared to NMPC.  

\subsubsection{Comparisons across multiple races}

\paragraph{Overtaking performance}  

\begin{figure}[htbp]
  \begin{minipage}[b]{\linewidth}
    \centering
    \includegraphics[width=0.6\linewidth]{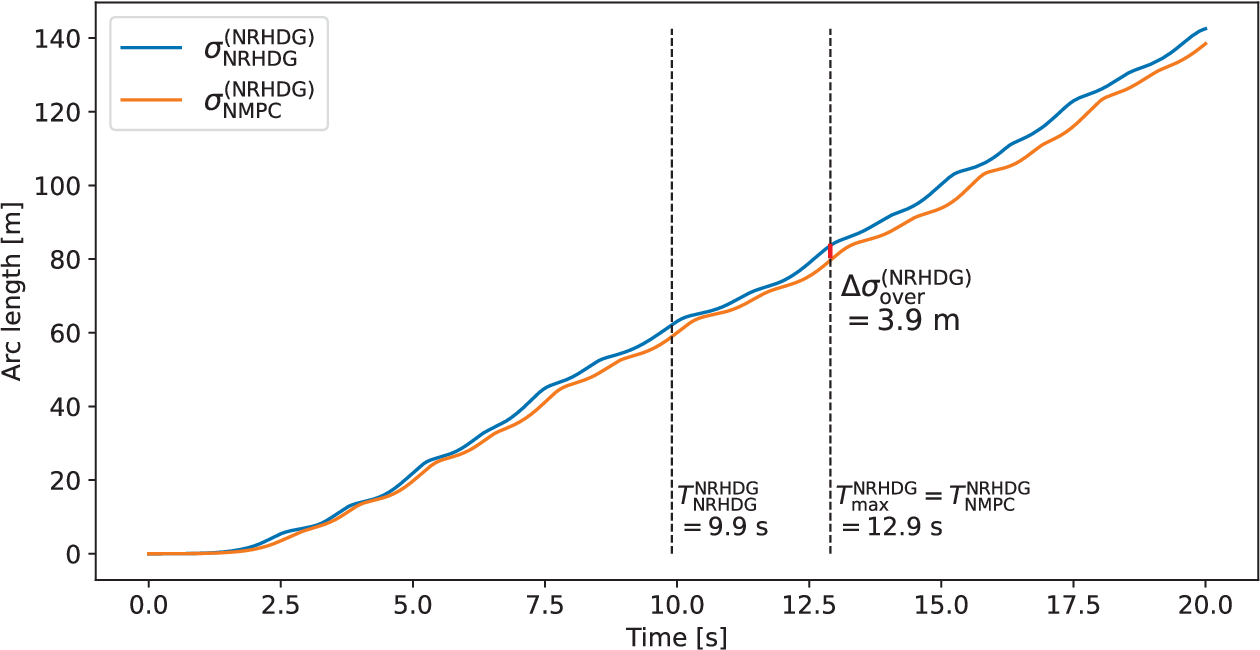}
    \subcaption{Comparison of $\text{Race}(\text{NRHDG},\text{NRHDG})$ and $\text{Race}(\text{NRHDG},\text{NMPC})$.}
    \label{fig:overtakeD}
  \end{minipage} \\
  \begin{minipage}[b]{\linewidth}
    \centering
    \includegraphics[width=0.6\linewidth]{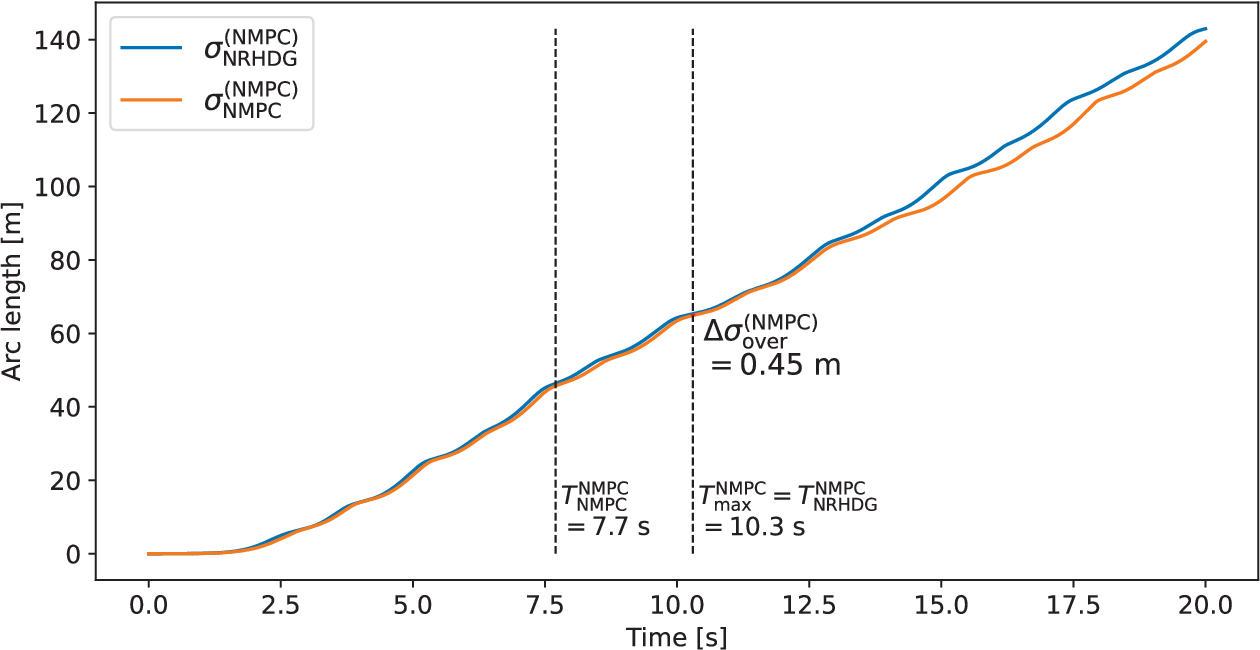}
    \subcaption{Comparison of $\text{Race}(\text{NMPC},\text{NRHDG})$ and $\text{Race}(\text{NMPC},\text{NMPC})$.}
    \label{fig:overtakeM}
  \end{minipage}
  \caption{Comparison of the rear-start drone's arc-length progress in $\text{Race}(A,\text{NRHDG})$ and $\text{Race}(A,\text{NMPC})$ for $A \in \{\text{NRHDG},\text{NMPC}\}$. 
  Vertical dashed lines indicate the overtaking times in the races. 
  NRHDG outperforms NMPC in overtaking performance for both choices of $A$, as indicated by $\Delta \sigma^{(A)}_{\text{over}} > 0$. }
  \label{fig:overtake}
\end{figure}

We evaluate the overtaking performance by comparing two races: $\text{Race}(A,\text{NRHDG})$ and $\text{Race}(A,\text{NMPC})$ with $A \in \{\text{NRHDG},\text{NMPC}\}$. 
The comparison of $\text{Race}(\text{NRHDG},\text{NRHDG})$ and $\text{Race}(\text{NRHDG},\text{NMPC})$ is shown in Figure \ref{fig:overtakeD} in terms of the progress of the drone starting from the rear position.
Two vertical dashed lines indicate the overtaking times in the two races, $T^\text{NRHDG}_\text{NRHDG} = 9.9$ s and $T^\text{NRHDG}_\text{NMPC} = 12.9$ s, respectively.  
Figure \ref{fig:overtakeD} shows the plots of 
$\sigma^{(\text{NRHDG})}_{\text{NRHDG}}(t)$ and $\sigma^{(\text{NRHDG})}_{\text{NMPC}}(t)$. 
Then, the overtaking performance difference $\Delta \sigma^{(\text{NRHDG})}_\text{over}$ is calculated to be $3.9$ m at the maximum overtaking time $T^\text{NRHDG}_\text{max} = T^\text{NRHDG}_\text{NMPC} = 12.9$ s, and 
$\Delta \sigma^{(\text{NRHDG})}_{\text{over}} > 0$ holds as expected. 
That is, NRHDG makes more progress than NMPC while overtaking the same opponent, namely, NRHDG. 
The comparison of $\text{Race}(\text{NMPC},\text{NRHDG})$ and $\text{Race}(\text{NMPC},\text{NMPC})$ is shown in Figure \ref{fig:overtakeM}. As shown in the figure, 
the overtaking times in the two races are $T^\text{NMPC}_\text{NMPC} = 7.7$ s and $T^\text{NMPC}_\text{NRHDG} = 10.3$ s, respectively. 
The overtaking performance difference 
$\Delta \sigma^{(\text{NMPC})}_\text{over}$ is calculated to be $0.45$ m at the maximum overtaking time $T^\text{NMPC}_\text{max} = T^\text{NMPC}_\text{NRHDG} = 10.3$ s, and 
$\Delta \sigma^{(\text{NMPC})}_{\text{over}} > 0$ holds. Therefore, NRHDG still makes more progress than NMPC while overtaking the same opponent, namely, NMPC. 

\paragraph{Obstructing performance}

\begin{figure}[htbp]
  \begin{minipage}[b]{\linewidth}
    \centering
    \includegraphics[width=0.6\linewidth]{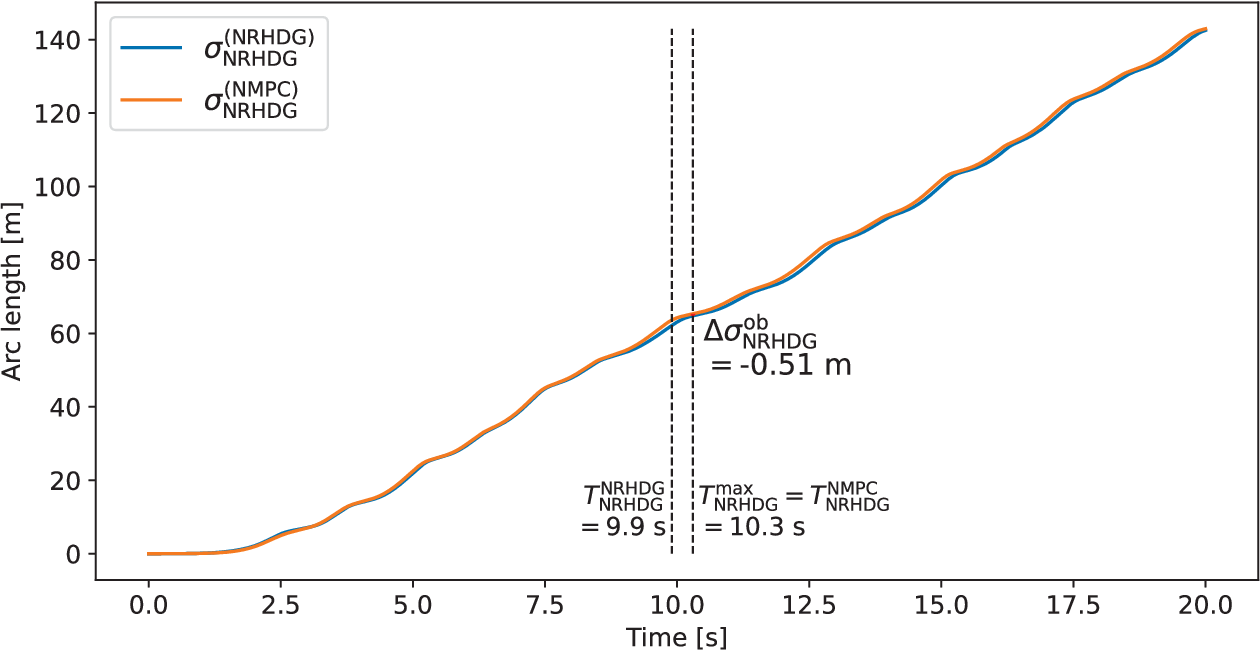}
    \subcaption{Comparison of $\text{Race}(\text{NRHDG},\text{NRHDG})$ and $\text{Race}(\text{NMPC},\text{NRHDG})$.}
    \label{fig:observeD}
  \end{minipage} \\
  \begin{minipage}[b]{\linewidth}
    \centering
    \includegraphics[width=0.6\linewidth]{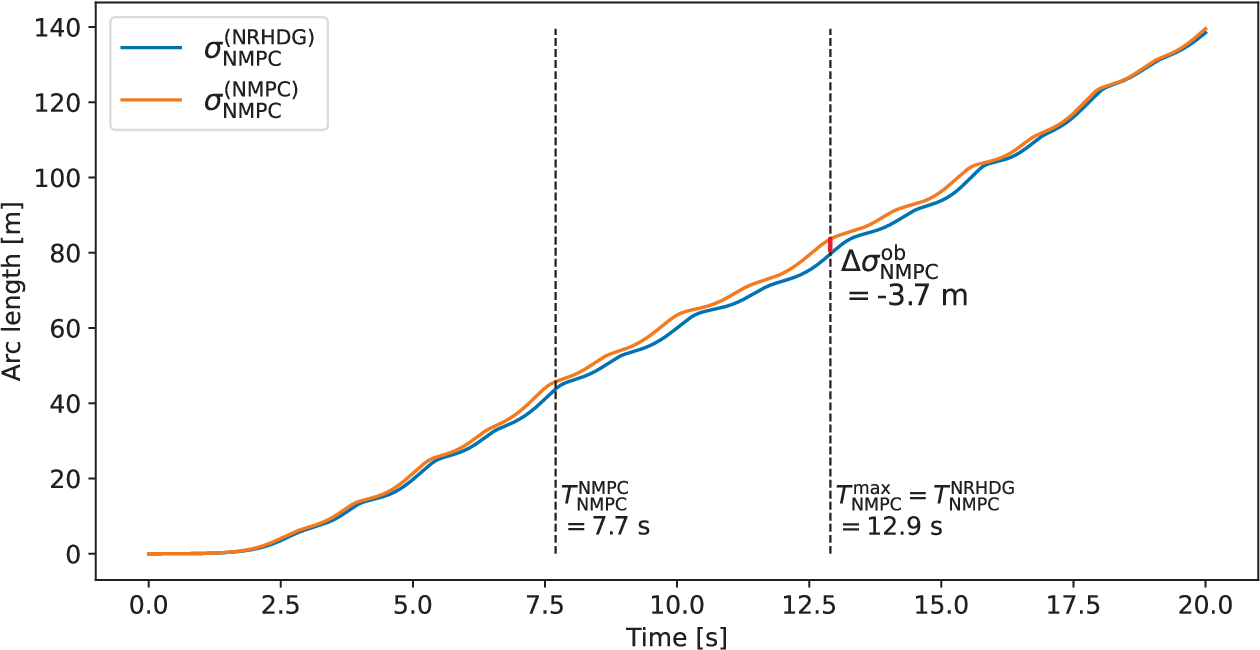}
    \subcaption{Comparison of $\text{Race}(\text{NRHDG},\text{NMPC})$ and $\text{Race}(\text{NMPC},\text{NMPC})$.}
    \label{fig:observeM}
  \end{minipage}
  \caption{Comparison of the rear-start drone's arc-length progress in $\text{Race}(\text{NRHDG},B)$ and $\text{Race}(\text{NMPC},B)$ for $B \in \{\text{NRHDG},\text{NMPC}\}$. 
  Vertical dashed lines indicate the overtaking times in the races. 
  NRHDG outperforms NMPC in obstructing performance for both choices of $B$, as indicated by $\Delta \sigma^{\text{ob}}_{B} < 0$. }
  \label{fig:observe}
\end{figure}

We also evaluate the obstructing performance by comparing two races: $\text{Race}(\text{NRHDG}, B)$ and $\text{Race}(\text{NMPC}, B)$ with $B \in \{\text{NRHDG}, \text{NMPC} \}$. The comparison of $\text{Race}(\text{NRHDG},\text{NRHDG})$ and $\text{Race}(\text{NMPC},\text{NRHDG})$ is shown in Figure \ref{fig:observeD}. The graph shows the progress of the rear-start NRHDG rather than that of the front-start obstructing drone. 
Figure \ref{fig:observeD} shows the plots of $\sigma^{(\text{NRHDG})}_{\text{NRHDG}}(t)$ and $\sigma^{(\text{NMPC})}_{\text{NRHDG}}(t)$ with overtaking times indicated by two vertical dashed lines, $T^\text{NRHDG}_\text{NRHDG} = 9.9$ s and $T^\text{NMPC}_\text{NRHDG} = 10.3$ s, respectively.
From the figure, the obstructing performance difference $\Delta \sigma^{\text{ob}}_{\text{NRHDG}} = -0.51$ m is calculated at the maximum overtaking time $T^\text{max}_\text{NRHDG} = T^\text{NMPC}_\text{NRHDG} = 10.3$ s, and $\Delta \sigma^{\text{ob}}_{\text{NRHDG}} < 0$ holds. Therefore, NRHDG obstructs the same opponent, NRHDG, more effectively than NMPC. 
The comparison of $\text{Race}(\text{NRHDG},\text{NMPC})$ and $\text{Race}(\text{NMPC},\text{NMPC})$ is shown in Figure \ref{fig:observeM}. The figure shows overtaking times, $T^\text{NMPC}_\text{NMPC} = 7.7$ s and $T^\text{NRHDG}_\text{NMPC} = 12.9$ s, respectively, indicated by two vertical dashed lines. 
The obstructing performance difference $\Delta \sigma^{\text{ob}}_{\text{NMPC}} = -3.7$ m is calculated at the maximum overtaking time $T^\text{max}_\text{NMPC} = T^\text{NRHDG}_\text{NMPC} = 12.9$ s, and $\Delta \sigma^{\text{ob}}_{\text{NMPC}} < 0$ holds. 
Therefore, NRHDG obstructs the same opponent, NMPC, more effectively than NMPC.

\subsubsection{Comparisons across different conditions}

\begin{figure}[htbp]
\hfil
  \begin{minipage}[b]{0.45\linewidth}
    \centering
    \includegraphics[width=0.9\linewidth]{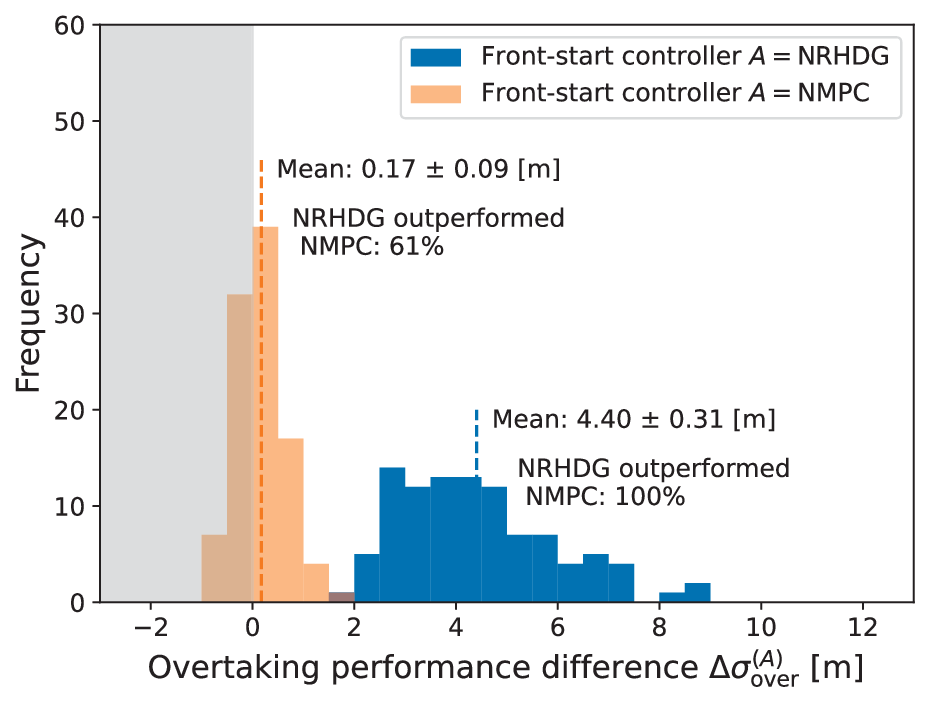}
    \subcaption{Overtaking performance difference between NRHDG and NMPC.}
    \label{fig:dsoverDandM_Ru2_20}
  \end{minipage} 
\hfil
  \begin{minipage}[b]{0.45\linewidth}
    \centering
    \includegraphics[width=0.9\linewidth]{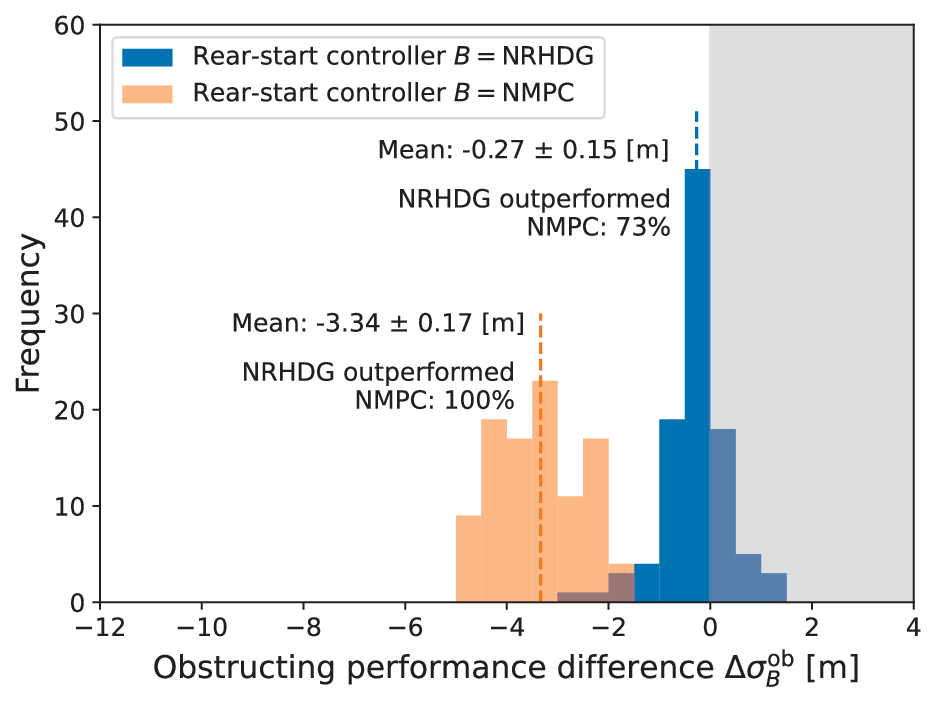}
    \subcaption{Obstructing performance difference between NRHDG and NMPC.}
    \label{fig:dsobDandM_Ru2_20}
  \end{minipage}
\hfil
  \caption{Histograms of performance differences between NRHDG and NMPC for randomly perturbed initial positions in the nominal case ($b=20$). Means are shown with $95\%$ confidence intervals. NRHDG does not outperform NMPC in the gray areas. }
  \label{fig:pd_Ru2_20}
\end{figure}

Furthermore, we conducted a set of simulations with 100 sets of randomly perturbed initial conditions. 
The initial positions of the two drones were perturbed by random vectors independently drawn from a uniform distribution over $[-1, 1]^3 \subset \mathbb{R}^3$. 
The same set of random initial conditions was used for the four race scenarios, $\text{Race}(A,B)$ for $A, B \in \{\text{NRHDG},\text{NMPC} \}$, resulting in a total of 400 simulations. 
The overtaking performance difference $\Delta \sigma^{(A)}_{\text{over}}$ in (\ref{eq:dsover}) and the obstructing performance difference $\Delta \sigma^{\text{ob}}_B$ in (\ref{eq:dsob}) were evaluated for the 100 cases against NRHDG and NMPC, respectively. 
Figure \ref{fig:pd_Ru2_20} shows the histograms of the overtaking performance differences $\Delta \sigma^{(A)}_{\text{over}}$ against a front-start controller $A \in \{\text{NRHDG},\text{NMPC} \}$ and obstructing performance differences $\Delta \sigma^{\text{ob}}_{B}$ against a rear-start controller $B \in \{\text{NRHDG},\text{NMPC} \}$, respectively, for the same setting as the previous simulations (nominal case) except for the initial conditions. 
The gray regions indicate ranges of performance-difference values for which NRHDG does not outperform NMPC. 
The means of the performance differences are shown with two-sided $95\%$ confidence intervals computed using the Student's $t$ distribution.
For the overtaking performance, $\Delta \sigma^{(\text{NRHDG})}_{\text{over}} > 0$ holds and, therefore, NRHDG outperforms NMPC when overtaking the front-start NRHDG in all 100 cases. 
When overtaking the front-start NMPC, NRHDG outperforms NMPC, i.e., $\Delta \sigma^{(\text{NMPC})}_{\text{over}} > 0$ holds, in 61\% of cases. 
Regarding the obstructing performance, NRHDG outperforms NMPC when obstructing NRHDG, i.e., $\Delta \sigma^{\text{ob}}_{\text{NRHDG}} < 0$ holds in 73\% of cases, and NRHDG outperforms NMPC when obstructing NMPC, i.e., $\Delta \sigma^{\text{ob}}_{\text{NMPC}} < 0$ holds in all 100 cases, respectively.

\begin{figure}[htbp]
\hfil
  \begin{minipage}[b]{0.45\linewidth}
    \centering
    \includegraphics[width=0.9\linewidth]{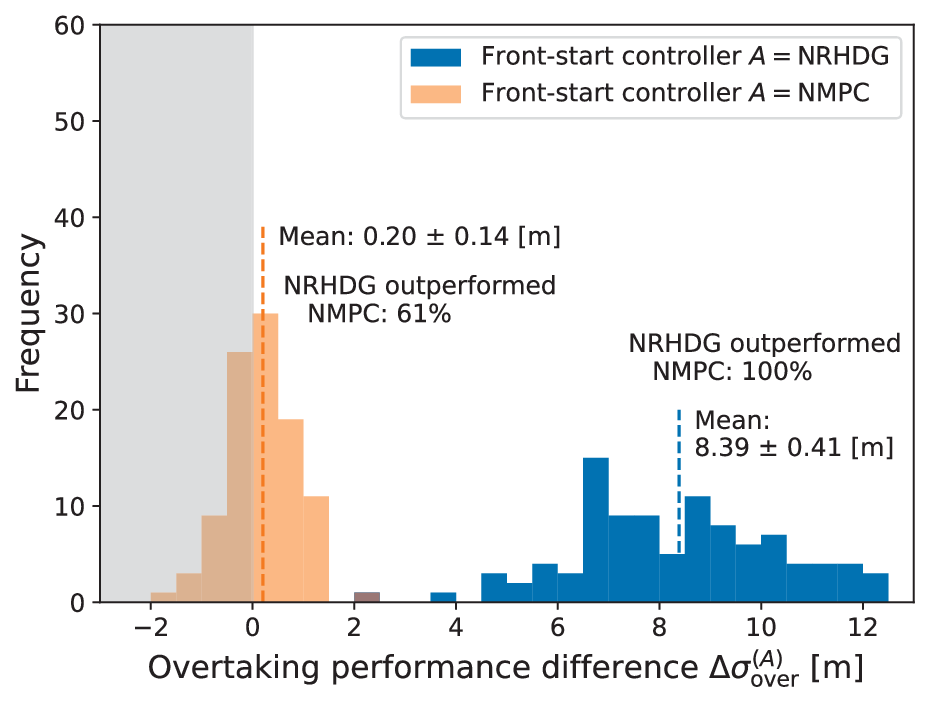}
    \subcaption{Overtaking performance difference between NRHDG and NMPC.}
    \label{fig:dsoverDandM_Ru2_25}
  \end{minipage} 
\hfil
  \begin{minipage}[b]{0.45\linewidth}
    \centering
    \includegraphics[width=0.9\linewidth]{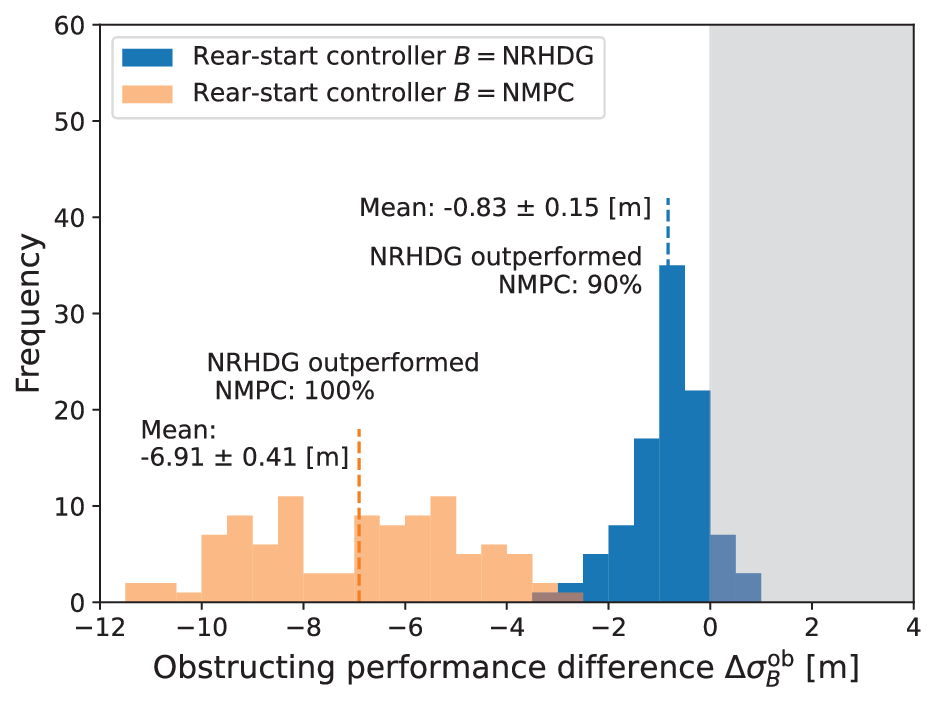}
    \subcaption{Obstructing performance difference between NRHDG and NMPC.}
    \label{fig:dsobDandM_Ru2_25}  \end{minipage}
\hfil
  \caption{Histograms of performance differences between NRHDG and NMPC for randomly perturbed initial positions with a slower rear-start drone ($b=25$) than the nominal case. Means are shown with $95\%$ confidence intervals. NRHDG does not outperform NMPC in the gray areas. }
  \label{fig:pd_Ru2_25}
\end{figure}

\begin{figure}[htbp]
\hfil
  \begin{minipage}[b]{0.45\linewidth}
    \centering
    \includegraphics[width=0.9\linewidth]{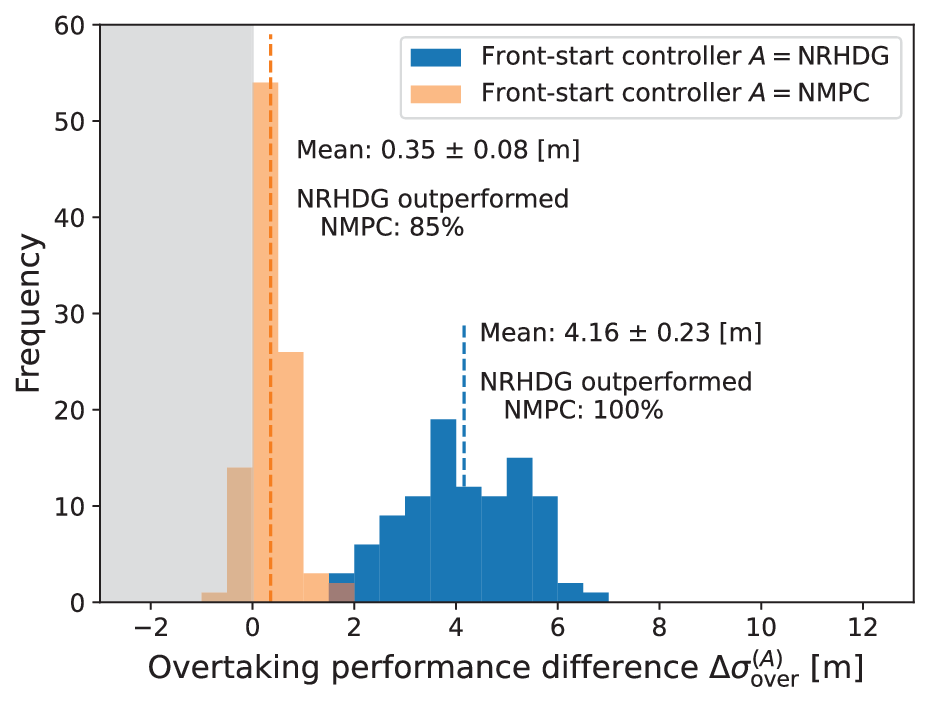}
    \subcaption{Overtaking performance difference between NRHDG and NMPC.}
    \label{fig:dsoverDandM_Ru2_15}
  \end{minipage} 
\hfil
  \begin{minipage}[b]{0.45\linewidth}
    \centering
    \includegraphics[width=0.9\linewidth]{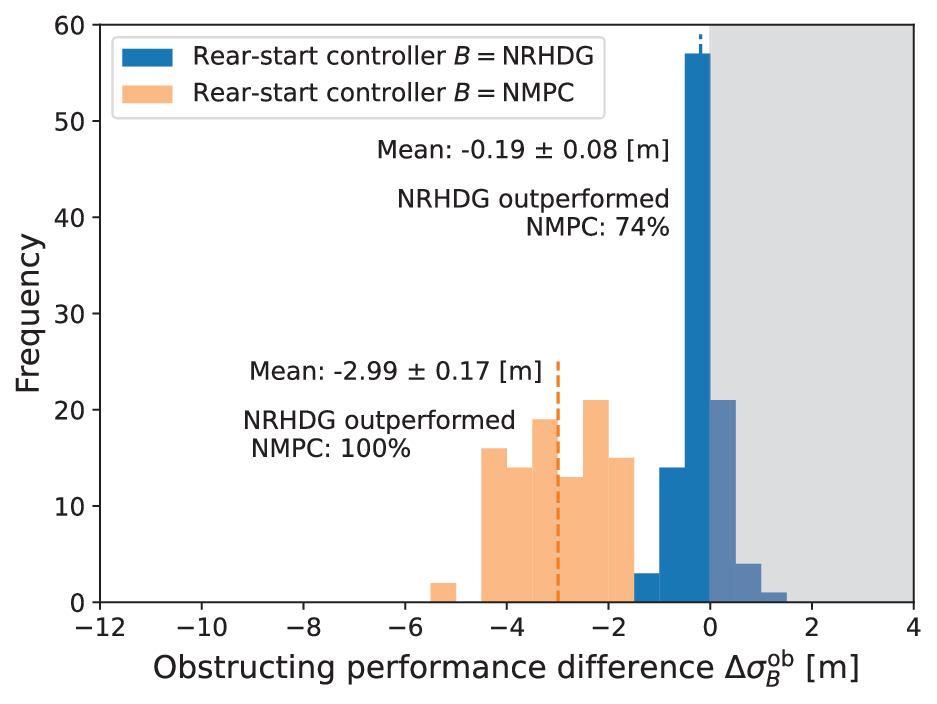}
    \subcaption{Obstructing performance difference between NRHDG and NMPC.}
    \label{fig:dsobDandM_Ru2_15}
  \end{minipage}
\hfil
  \caption{Histograms of performance differences between NRHDG and NMPC for randomly perturbed initial positions with a faster rear-start drone ($b=15$) than the nominal case. Means are shown with $95\%$ confidence intervals. NRHDG does not outperform NMPC in the gray areas. }
  \label{fig:pd_Ru2_15}
\end{figure}

To assess the influence of the relative speed capability on the differences in the controller's performance, we conducted additional simulations with the same set of random initial conditions as in the nominal case but with different input weights for the rear-start controller. 
Figure \ref{fig:pd_Ru2_25} shows performance differences for cases with a larger input weight ($b = 25$) for the rear-start drone, which corresponds to a smaller speed advantage of the rear-start drone than the nominal case. 
For the overtaking performance, NRHDG outperforms NMPC when overtaking the front-start NRHDG, i.e., $\Delta \sigma^{(\text{NRHDG})}_{\text{over}} > 0$ holds, in all 100 cases. 
When overtaking the front-start NMPC, NRHDG outperforms NMPC, i.e., $\Delta \sigma^{(\text{NMPC})}_{\text{over}} > 0$ holds, in 61\% of cases. 
Regarding the obstructing performance, NRHDG outperforms NMPC when obstructing NRHDG in 90\% of cases and in all 100 cases when obstructing NMPC, respectively. 
Figure \ref{fig:pd_Ru2_15} shows performance differences for cases with a smaller input weight ($b = 15$) for the rear-start drone, which corresponds to a larger speed advantage of the rear-start drone than the nominal case. 
For the overtaking performance, NRHDG outperforms NMPC in all 100 cases when overtaking the front-start NRHDG. 
When overtaking the front-start NMPC, NRHDG outperforms NMPC in 85\% of cases. 
Regarding the obstructing performance, NRHDG outperforms NMPC in 74\% of cases when obstructing NRHDG and in all 100 cases when obstructing NMPC, respectively.

Across all three conditions shown in Figures \ref{fig:pd_Ru2_20}--\ref{fig:pd_Ru2_15}, the 95\% confidence intervals of mean performance differences excluded zero for both overtaking and obstructing comparisons, indicating statistically significant superiority of NRHDG over NMPC under the uniform distribution of initial conditions.

\subsection{Discussion}

In the simulation results (Figs.\ \ref{fig:pd_Ru2_20}--\ref{fig:pd_Ru2_15}), NRHDG outperforms NMPC in all simulated cases when overtaking a front-start drone controlled by NRHDG with various initial conditions and different levels of speed advantage of the rear-start drone. 
This observation is consistent with the defining property of a saddle-point solution in a DGP: any unilateral deviation from the saddle-point solution cannot improve the deviating player's benefit. 
Moreover, when the front-start controller is NRHDG, the overtaking performance difference increases as the rear-start drone's input weight $b$ increases, that is, as its speed advantage decreases. 
This result indicates that NRHDG leverages a small speed advantage more effectively than NMPC does.

However, the advantage of NRHDG over NMPC is less pronounced when overtaking NMPC. 
This observation suggests that the difference in overtaking performance is smaller when overtaking a controller that does not explicitly model adversarial behavior. 
A potential limitation is that NRHDG may become overly conservative due to worst-case opponent modeling, especially when the opponent does not behave in a fully adversarial manner. 
Nevertheless, NRHDG still outperforms NMPC in at least 61\% of the cases when overtaking NMPC under different conditions. 
Moreover, the lower bounds of the 95\% confidence intervals of the mean overtaking performance differences remain positive across all race scenarios and the three levels of speed advantage. 
Therefore, we can conclude that NRHDG has better overtaking performance than NMPC.

The simulation results also show that NRHDG outperforms NMPC in all simulated cases when obstructing NMPC with various initial conditions and different levels of speed advantage for the rear-start drone. 
This can be interpreted as a consequence of the min-max structure in NRHDG, which explicitly penalizes the opponent's progress, whereas NMPC optimizes only its own objective. 
Moreover, when the rear-start controller is NMPC, the obstructing performance difference becomes more negative as the rear-start drone's input weight $b$ increases, that is, as its speed advantage decreases. 
This observation suggests that NRHDG more effectively obstructs a rear-start drone than NMPC does when the speed difference between the front-start and rear-start drones is small. 

However, the superiority of NRHDG over NMPC is less pronounced when obstructing NRHDG.  
This observation suggests that the rear-start NRHDG overestimates the adversarial obstructing behavior of the front-start NMPC and takes a conservative strategy in some cases. 
Nevertheless, NRHDG still outperforms NMPC in at least 73\% of the cases across all tested conditions. 
Moreover, the upper bounds of the 95\% confidence intervals of the mean obstructing performance differences remain negative across all race scenarios and the three levels of speed advantage. 
Therefore, we can conclude that NRHDG has better obstructing performance than NMPC.

Overall, these results provide empirical evidence that explicitly modeling adversarial interactions through NRHDG generally improves performance over NMPC in both overtaking and obstructing tasks under a wide range of conditions. 
A theoretical and quantitative characterization of the performance differences between controllers under various race conditions remains an important direction for future work.

\section{Conclusions} 

This study presents NRHDG, a game-theoretic control framework for competitive drone racing that addresses both path-following control and adversarial interactions. 
Building on a unified path-following formulation based on projection-point dynamics, the proposed approach eliminates the need for iterative distance minimization or projection approximation during online control. The proposed potential function further enables drones to switch between overtaking and obstructing strategies, while new performance metrics systematically evaluate overtaking and obstructing capabilities. 
Overall, the simulation results provide evidence that the proposed NRHDG framework can effectively capture competitive interactions in drone racing scenarios and generally achieves better performance than the baseline NMPC. The results also highlight the importance of explicitly modeling adversarial behavior when designing control strategies for competitive multi-agent systems. 
Specifically, for randomly generated initial conditions and different levels of speed advantage for the rear-start drone, the 95\% confidence intervals for the mean performance differences excluded zero, indicating statistically significant advantages of NRHDG over NMPC. 
Beyond drone racing, the developed principles and techniques have potential applications in other domains involving dynamic multi-agent interactions, such as autonomous vehicle coordination, robotic swarm navigation, and air traffic management. 
These applications highlight the broader significance of NRHDG for advancing control methodologies in competitive and dynamic systems.

Future research directions beyond the scope of the present study include adapting NRHDG to more complex racing environments with gates. 
Another possible extension is to consider races involving three or more drones, for which a multiplayer non-zero-sum game framework is necessary. 
Addressing uncertainties in drone dynamics and opponent strategies will also be critical for real-world implementation. This includes developing robust methods to handle unknown disturbances such as wind and measurement noise from sensors, and designing predictive models that account for stochastic behavior in opponents.



\bmsection*{Acknowledgments}
The authors thank Hiroshi Okajima for valuable comments on the draft of this paper.


\bmsection*{Conflict of interest}
The authors declare no conflict of interest.

\bmsection*{Data Availability Statement}
AutoGenU for Jupyter is available at \url{https://github.com/ohtsukalab/autogenu-jupyter}. 
Other data that support the findings of this study are also available from the corresponding author, Toshiyuki Ohtsuka, upon reasonable request. 

\bibliography{drone_racing}






\end{document}